\documentclass[twocolumns]{aa}  

\usepackage{graphicx}
\usepackage{subfig}
\usepackage{enumitem}
\usepackage{txfonts}
\usepackage{hyperref}
\usepackage{mathtools}

\def\tout{$\theta_{out}$}
\def\tin{$\theta_{in}$}
\def\i{$i$}
\def\r0{$r_0$}
\def\xstar{{\it XSTAR}}

\begin{document} 

\title{Modelling absorption and emission profiles from accretion disc winds with WINE}
\author{
A. Luminari\inst{1,2}, E. Piconcelli\inst{2}, F. Tombesi\inst{3,2,4} F. Nicastro\inst{2}, F. Fiore\inst{5,6}
}

\institute{
INAF - Istituto di Astrofisica e Planetologia Spaziali, Via del Fosso del Caveliere 100, I-00133 Roma, Italy
\and
INAF - Osservatorio Astronomico di Roma, Via Frascati 33, 00078 Monteporzio, Italy
\and
Dept. of Physics, University of Rome “Tor Vergata”, Via della Ricerca Scientifica 1, 00133 Rome, Italy
\and
INFN – Roma Tor Vergata, Via Della Ricerca Scientifica 1, 00133 Rome, Italy
\and
INAF – Osservatorio Astronomico di Trieste, Via G. B. Tiepolo 11, 34143 Trieste, Italy
\and
IFPU – Institute for Fundamental Physics of the Universe, Via Beirut 2, 34014 Trieste, Italy
}
\date{Received 16 January 2024; accepted 10 October 2024}

\abstract
{Fast and massive winds are ubiquitously observed in the UV and X-ray spectra of Active Galactic Nuclei (AGN) and other accretion-powered sources. Several theoretical and observational pieces of evidence suggest they are launched at accretion disc scales, carrying significant mass and angular momentum. Thanks to such high energy output, they may play an important role in transferring the energy released by accretion to the surrounding environment. In the case of AGNs, this process can help setting the so-called coevolution between the AGN and its host galaxy, which mutually regulates their growth across cosmic times. To precisely assess the effective role of UV and X-ray winds at accretion disc scales, it is necessary to accurately measure their properties, including mass and energy rates. However, this is a challenging task both due to the limited signal to noise ratio of available observations and to limitations of the  models currently used in the spectral analysis.}
{We aim to maximise the scientific return of current and future observations by improving the theoretical modelling of these winds through our Winds in the Ionised Nuclear Environment (WINE) model. WINE is a spectroscopic model specifically designed for disc winds in AGNs and compact accreting sources, which couples photoionisation and radiative transfer with special relativistic effects and a three-dimensional modellisation of the emission profiles.}
{We explore with WINE the main spectral features associated to the disc winds in AGNs, with particular emphasis on the detectability of the wind emission in the total transmitted spectrum. We explore the impact of the wind ionisation, column density, velocity field and geometry in shaping the emission profiles. We simulate observations with the X-ray microcalorimeters \textit{Resolve} onboard the recently-launched \textit{XRISM} satellite and the \textit{X-IFU} onboard the future \textit{Athena} mission. This allows us to assess the capabilities of these telescopes for the study of disc winds in X-ray spectra of AGNs for the typical physical properties and exposure times of the sources included in the \textit{XRISM} Performance Verification phase.}
{The wind kinematic and geometry (together with the ionisation and column density) deeply affect both shape and strength of the wind spectral features. Thanks to this, both \textit{Resolve} and, on a longer timescale, \textit{X-IFU} will be able to accurately constrain the main properties of disc winds in a broad range of ionisation, column density and covering factor. We also investigate the impact of the Spectral Energy Distribution (SED) on the resulting appearance of the wind. Our findings reveal a dramatic difference in the gas opacity when using a soft, Narrow Line Seyfert 1-like SED compared to a canonical powerlaw SED with spectral index $\Gamma \approx 2$.}
{}


\titlerunning{Modelling disc winds with WINE}
\authorrunning{A. Luminari et al.}
\maketitle

\section{Introduction}
\label{intro}
Disc winds are ubiquitously observed in many accreting sources, from compact sources to stellar and super massive black holes at the centre of Active Galactic Nuclei (AGNs). AGN-driven winds are considered as one of the fundamental mechanisms in shaping the accretion process itself and the interaction with the surrounding host environment. Indeed, for AGNs, winds are regarded as one of the key player for the feedback towards the host galaxy (see e.g. \citealp{Faucher12,King15,Fiore17,torrey20} and references therein).

Given the high degree of ionisation displayed by the gas, most of the observable transitions fall in the UV and X-ray bands. By comparing the observed spectra with simulated ones it is possible to constrain the wind properties, first of all $N_H, v_{out}, \xi$, i.e. the column density, outflow velocity and ionisation degree. The latter quantity is defined as the ratio between the ionising luminosity $L_{ion}$ (i.e., the luminosity above the ionisation threshold of Hydrogen, E=13.6 eV) and the product between gas density and distance:  $\xi=L_{ion}/n r^2$ \cite{tarter69}. In the UV range, most of the observed features are classified as Broad Absorption Lines (BAL), traced by mildly-ionised lines such as CIV, NV, Si IV with velocities from thousand km s$^{-1}$ up to 0.3 c, being c the light speed (see e.g. \citealp{murray95,arav01,bruni19,bischetti22,vietri22}). In the X-ray band the two broad classes into which winds are commonly classified are Warm Absorbers (WAs, observed in $\sim$ 50\% of the AGNs, \citealp{blustin05,pico05}), where the ionisation status is roughly consistent with the ions giving rise to BALs, i.e. $-1 \lesssim \log(\xi / erg\ cm\ s^{-1}) \lesssim 3$ and $N_H \lesssim 10^{23} cm^{-2}$, and higher-ionisation, mildly relativistic Ultra-Fast Outflows (UFOs), with $\log(\xi / erg\ cm\ s^{-1}) \gtrsim 3, N_H > 10^{23} cm^{-2}$ and $v_{out}$ around 0.1 c, with a high-velocity tail until 0.3 c (see e.g. \citealp{pounds03,cappi09,Fiore17,laha21,lne21,chartas21,matzeu23} and refs. therein). It is important to note that, being most of these winds spatially unresolved, line-of-sight integrated spectroscopy provides the prime tool for their study.

From the above spectral quantities, estimates of the wind energetic, namely the mass and energy fluxes, are usually estimated under the assumption of spherical symmetry and classic (i.e. non relativistic) dynamics as (see e.g. \citealp{crenshaw2012,tcr12}):
\begin{equation}
\begin{split}
& \Dot{M}_{out}=4 \pi r_0 N_H \mu m_p C_f v_{out} \\
& \Dot{E}_{out}= \frac{1}{2} \Dot{M}_{out} v_{out}^2
\end{split}
\end{equation}
where $r_0, C_f$ are the wind launching radius and covering factor, respectively, and $\mu, m_p$ are the mean atomic mass per proton and the proton mass (see \citealp{kne07,fiore23} for alternative formulations).

Several photoionisation codes are available to fit the observed spectra, such as Cloudy \citep{Cloudy17}, XSTAR \citep{kallman21} and SPEX \citep{spex}. For given incident Spectral Energy Distribution (SED), $N_H, \xi$, they allow to compute the ionisation balance of the gas and obtain both their transmitted and emitted spectra, which can then be fitted to the observations. These codes allow an extensive exploration of the parameter space of disc winds and robust estimates of their energetic, thus readily allowing to assess the wind role and dynamics. However, it must be noted that such codes are thought to be as much universal as possible, allowing to reproduce virtually any astrophysical setting, from the atmosphere of stars to diffuse Ly$\alpha$ nebulae. Their physical picture is not therefore able to account for the full complexity of the disc wind phenomenon, which shows peculiar properties, such as strong velocity gradients, mildly relativistic velocities and a non-spherical geometry, strongly deviating from the spherical symmetry. Such features can largely affect the gas spectroscopic appearance and, thus, must be properly accounted for to obtain reliable synthetic wind spectra and to constrain the wind properties meaningfully. Specifically, we note three major assumptions of present photoionisation codes that are not met by disc winds:
\begin{itemize}
    \item The gas is at rest with respect to the luminosity source (i.e. zero net velocity)
    \item The thermal motion of the gas regulates the line broadening (i.e., gas velocity shearing is negligible)
    \item The gas cloud is spherically symmetric around the radiation source, resulting in emission profiles with Gaussian shapes\footnote{We note that the above codes account, to different extents, for the covering factor of the gas and the presence of velocity gradients. However, such properties only affect the ionisation balance and the radiative transfer but do not affect the profile of the spectral lines.}
\end{itemize}
Together with the signal-to-noise and resolving power limitations of current X-ray spectra, such theoretical assumptions often prevent a reliable estimate of both the wind covering fraction $C_f$ (which is usually assumed by looking at the statistical recurrence of outflows in large samples) and $r_0$, which is estimated through indirect arguments, i.e. by equating $v_{out}$ to the escape velocity or through the definition of $\xi$ and assuming a constant wind density and a plane-parallel geometry (see e.g. \citealp{tcr11} and discussion in \citealp{llt21}). As a result, the derived $\Dot{M}_{out}, \Dot{E}_{out}$ usually have order-of-magnitude uncertainties, which prevent a detailed assessment of the feedback of such outflows on both the accretion process and the host environment (see e.g. Figures 2 and 3 in \citealp{tcr12} and Figure 8 in \citealp{smith19}).

Of particular interest are the so-called P-Cygni profiles, where, in analogy with the stellar atmospheres \citep{castor79}, a line transition is detected both in emission and absorption, with the latter at slightly blueshifted (i.e. higher) energies due to the bulk motion of the absorbing gas along the line of sight (LOS).
Under the hypothesis that these spectral components are produced in the same expanding gas, such remarkable spectral feature allows to probe both the outflowing gas along our LOS and its distribution over the solid angle. This clearly represents a great advantage for measuring physical and dynamical properties of the gas with respect to spectra where only absorption lines are detected.
Notable examples of P-Cygni profiles associated to UFOs in AGN have been discovered in the X-ray spectra of PDS 456 \citep{nrg15,lpt18}, 1H 0707-495 \citep{hagino16}, PG1448+273 \citep{kosec20,llt21}, I Zwicky 1 \citep{reeves19}. However, in most of the observations it is difficult to statistically rule out a non-negligible contribution by the accretion disc reflection to the emission profile (see e.g. \citealp{ppf17,middei23}), as discussed in detail by \cite{parker22} through sets of dedicated X-ray simulations. Updated, more accurate disc wind models are therefore needed to shed light on this intriguing aspect of disc winds and the interplay of their spectroscopy imprint with further emission components not related to the wind, especially given the unprecedented resolving power offered by the microcalorimeter {\it Resolve} onboard {\it XRISM} \citep{tashiro20} and the {\it Athena}'s X-IFU \citep{didier19,xifu22}.

To address the limitations discussed above and get a more accurate description of disc winds we developed the Wind in the Ionised Nuclear Environment (WINE) spectroscopic model. WINE features accurate photoionisation balance and radiative transfer through the XSTAR code and a detailed modelling of the dynamic and geometry of disc winds from compact sources and accreting black holes. Its main features are the inclusion of special relativity in both gas absorption and emission, a detailed treatment of the wind density and velocity profiles and a proper representation of the wind geometry, in order to deal with the assumptions listed above. The result is a physically motivated modelling of the wind emission and absorption profiles, that allows to constrain self-consistently the main wind properties and energetic. 
A description of the Monte Carlo approach is outlined in \citep{lpt18} (where it is also applied to the X-ray UFO in the quasar PDS 456), while special relativity effects are discussed in \cite{ltp20,marzi23} and applications of WINE to the UFOs in the AGNs PG1448+273 and NGC 2992 are presented in \cite{llt21,luminari23a}, respectively.

The structure of this paper is as follows. In Sect. \ref{scheme} we illustrate the structure of WINE and its free parameters, and we show characteristic WINE-generated spectra in Sect. \ref{outputs}. Then, we investigate the detectability of the emission profiles for typical UFO parameters, both through an algorithm which scans the wind spectra (Sect. \ref{pcyg}) and through simulated observations both with the {\it Resolve} microcalorimeter onboard {\it XRISM} (Sect. \ref{resolve-sim}) and the \textit{Athena X-IFU} (Sect. \ref{athena-sims}). Finally, we discuss and summarise our results in Sect. \ref{discuss}.

Note that in the following we will focus on the typical parameter range of Warm Absorbers and UFOs in AGNs, but WINE is perfectly fit for the study of any kind of disc wind arising from compact sources. Hereafter, $\xi, N_H$ will be expressed in units of $erg\ cm\ s^{-1}$ and $cm^{-2}$, respectively, and we omit them for ease of reading.

\section{The working scheme of WINE}
\label{scheme}
We designed WINE as a self-consistent model for disc winds, able to directly probe the main properties of the intervening gas - ionisation, velocity, column density, radial location, geometry - and, thus, to reliably estimate its energetic. We tuned the number of free parameters to the (expected) constraining power of the microcalorimeter {\it Resolve} onboard {\it XRISM} satellite, but we will also discuss possible avenues for further improvements.

For a given geometry and dynamics of the wind (which are set by the input parameters), the gas column density is sliced in a series of geometrically thin shells. Photoionisation computation starts from the innermost shell and is then propagated outwards. The resulting quantities - transmitted spectrum and gas emissivity - are processed to compute the wind absorption and emission profiles as a function of the physical properties of the wind (velocity, opening angles, line of sight, etc.). As we will discuss in detail below, a number of input parameters regulate the ionisation structure of the gas, while others regulate the kinematics and the geometry and, thus, the observational appearance of the gas spectral features.

\subsection{Wind geometry and dynamics}
\begin{figure}
\centering
\includegraphics[width=\columnwidth]{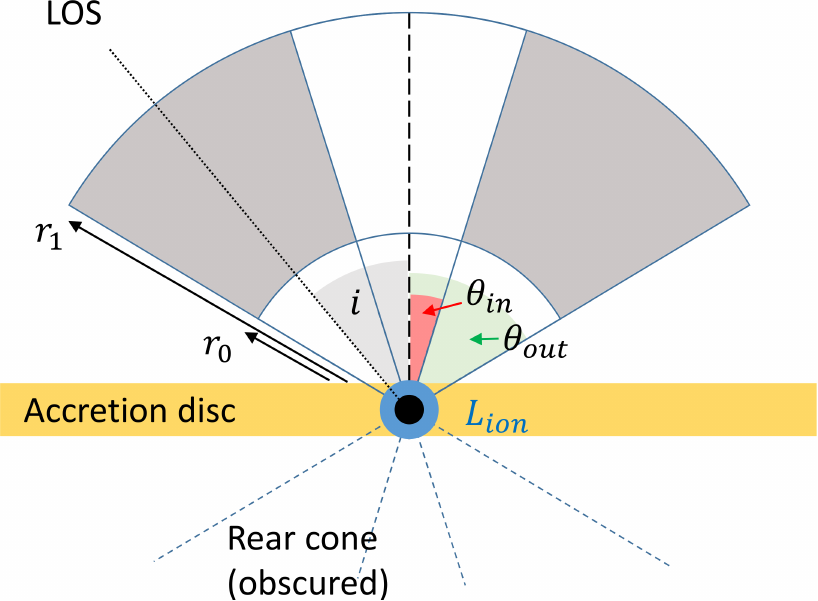}
\caption{Geometry of the WINE model. Grey shaded area indicate the wind volume. $\theta_{in}, \theta_{out}$ represents the wind inner and outer opening angle with respect to the symmetry axis (vertical dotted line). The line of sight (LOS) has an inclination angle $i$ with respect to the axis. Accretion disc is represented with a yellow bar, and the black and blue dots indicate respectively the accreting and the luminosity sources, assumed as point like. Finally, $r_0, r_1$ indicate the wind initial and final radius, respectively.}
\label{4cone}
\end{figure}

We assume a biconical geometry for the wind (see Fig. \ref{4cone}), centred on the accreting source (represented as a black dot) and with the same symmetry axis of the accretion disc (assumed to be planar). We assume the wind to be directed radially outward (but see Appendix \ref{appendix-vrot} for further discussion on this) and we consider that the rear cone is obscured by the disc (in yellow). The luminosity source is point like and located in the centre (blue region in Fig. \ref{4cone}). The wind is enclosed by an initial and final radius, $r_0, r_1$, and has an inner and outer opening angle, \tin\ and \tout, respectively (shown in red and green in Fig. \ref{4cone}), while the inclination of the line of sight (LOS) \i\ is shown in light grey.
The geometry is in line with most of the observations and the models of accretion disc wind in the literature \citep{psk00,pk04,kne07,omm09,fkc10,hagino15,matthews16a,maksym23}. Moreover, galactic-scale outflows also show quasi-spherical or biconical morphologies \citep{rupke13,feruglio15,venturi18,mingozzi19,Menci19,menci23}.

According to this geometry, the wind covering factor $C_f$ is:
\begin{equation}
C_f=\cos{\theta_{in}}-\cos{\theta_{out}}
\label{cf_eq}
\end{equation}
The inner cavity, specified by \tin, can also be interpreted as a first-order parametrisation for (possible) density variations along the polar angle. Such variations are indeed expected for both radiative and magnetohydrodinamic (MHD) accelerations, which produce variable wind distributions along the equatorial and the poloidal directions. As an example for magnetically-driven winds, the interplay between poloidal and toroidal magnetic fields may lead to a collimated, relativistic jet-like outflow around the rotation axis and a smooth, mildly relativistic outflow outside the jet region \citep{fkc10,fukumura14,yuan15,Cui20}. In this case, we expect the jet region to lie in the $0< \theta < \theta_{in}$ interval, and the wind in the $\theta_{in} < \theta < \theta_{out}$ one. 
Simulations of radiatively driven winds, instead, suggest that most of the wind develops closer to the equatorial region, while the polar one is associated with failed winds due to the overionisation by the central luminosity source \citep{pk04,higginbottom14}.
Moreover, a configuration with $\theta_{in} \approx \theta_{out}$ is able to reproduce a funnel-shaped wind; such structure is foreseen by several theoretical models (see e.g. \citealp{elvis00,matthews16,matthews20,sim08,sim10a,sim10b}) with the aim to unify most of the outflowing structures observed in the optical/UV and X-ray bands, i.e. from the Broad and the Narrow Line Regions up to UFOs.

The gas density and velocity are parametrised as power-law functions of the radial coordinate of the cone $r$ as follows:
\begin{equation}
n(r)=n_0 \cdot \bigg( \frac{r_0}{r} \bigg)^{\alpha}
\label{WINE:n_r}
\end{equation}
\begin{equation}
v(r)=v_0 \cdot \bigg( \frac{r_0}{r} \bigg)^{\alpha_v}
\label{WINE:v_r}
\end{equation}
where $n_0\equiv n(r=r_0), v_0\equiv v(r=r_0)$. Several analytical calculations point toward such power-law trends (see e.g. \citealp{Faucher12,Menci19}) in the case of a fast wind expanding in a perturbation-free environment. Specifically, in the case of momentum- and energy-conserving flows the following relations hold \citep{Faucher12}:
\begin{equation}
\alpha_v = \begin{cases} (2-\alpha)/2 & \mbox{Momentum conserving} \\ (2-\alpha)/3 & \mbox{Energy conserving} \end{cases}
\label{beta}
\end{equation}
so it is possible to link the velocity profile to the density one. We focus on an "ideal" (perturbation-free) outflow since we expect strong inhomogeneities to be suppressed at typical disc wind scales. Even though an intrinsic degree of anisotropy in the gas properties - density, temperature - is naturally expected at the outflow launching point, the strong acceleration gradients and the extreme velocities, which make the gas trans- to super-sonic (see e.g. \citealp{psk00,Cui20,yama21}), are expected to quench the growth of physical gradients and the setting of multi-phase flows. The development of a fully inhomogeneus flow, i.e. $\delta \rho / \langle \rho \rangle \gtrsim 1 , \delta T / \langle T \rangle \gtrsim 1 $ (where $\delta \rho = | \rho - \langle \rho \rangle | $ and $\langle \rho \rangle$ is the average density and the same holds for the temperature $T$) is typically expected at WA and BLR scales. The theoretical picture, however, is not yet fully established, and we refer to \cite{dannen20,waters21,waters22} and references therein for further discussions. 

We perform a first-order radiative transfer to account for the absorption of the gas emission spectrum by the gas column itself. As discussed in Sect. \ref{em_profiles}, such absorption is negligible for typical UFO ionisation degrees and velocities, but it becomes appreciable for higher gas opacity, i.e. low ionisation and higher column densities, $log(\xi) \leq 3, N_H \geq 10^{23}$.

\subsection{Input parameters}
The parameters regulating photoionisation and radiative transfer (via the \textit{XSTAR} code) are the following:
\begin{itemize}
\item $L_{ion}, S_I$, the incident luminosity and spectrum, respectively (either tabulated or analytic).
\item $\xi_0$, the ionisation parameter at the inner face of the gas cloud
\item $v_0, \alpha_v$, the starting velocity and the index in Eq. \ref{WINE:v_r}. $\alpha_v$ can also be linked to $\alpha$ using the relations of Eq. \ref{beta}.
\item $N_H^{max}$, the total column density of the wind, and the linear or logarithmic stepping pace
\item \r0\ , the initial radius of the wind. 
\item $\alpha$, the index of the density profile in Eq. \ref{WINE:n_r}
\item $v_{turb}$, the gas turbulent velocity. This can be either set by the user directly or computed self-consistently by the code following Eq. \ref{vturb} below. 
\end{itemize}
Metallicity is a free parameter and hereafter we adopt the standard solar values of Sanders, Grevesse, Noel (1968), i.e. the default XSTAR ones. Once photoionisation computation is performed, the transmitted spectrum and the line emissivities are used to obtain the wind absorption and emission spectrum. The line profiles are a function of the wind geometry and dynamics, which is set by the following free parameters (see Fig. \ref{4cone}): 
\begin{itemize}
\item $v_0, \alpha_v$, i.e. the velocity profile (see above)
\item $r_0$, the launching radius (see above)
\item $\theta_{out}$, the opening angle of the wind cone
\item $i$, the inclination of the line of sight with respect to the symmetry axis of the cone
\item \tin, the opening angle of the inner wind cavity
\end{itemize}
For practical purposes, when computing tables of absorption and emission spectra it is possible to let WINE iterate over ranges of input values. 

\subsection{Ionisation and radiative transfer}
\label{abs}
The wind ionisation and radiative transfer is computed in WINE through a multilayer approach with repeated calls to a ionisation code, similarly to what already done in previous works, e.g. \citet{schurch07,fkc10,saez11}. For a given $N_H$, the wind is divided in $M$ linear(or log)-spaced slabs, each one with a column density $\delta N_H^m$. For a given set of input parameters, photoionisation computation is started from the first shell using XSTAR and then propagated outward up to the $M$-th shell. This slicing allows to implement the wind velocity and density profiles of Eqs. \ref{WINE:n_r}, \ref{WINE:v_r} and to build absorption and emission templates for increasing $N_H$. The shell thickness $\delta N_H^m$ (or, equivalently, $M$) is a free parameter and has to be tuned so that the variation of $n, r, log(\xi)$ within each shell is negligible. In all the cases presented in this paper, we find that $\delta N_H^m=10^{23} cm^{-2}$ allows to optimally sample the wind column.

For a given $\xi_0, r_0$, the initial density of the wind $n_0$ is determined by inverting the definition of $\xi_0$:
\begin{equation}
\xi_0\equiv \xi(r=r_0)= \frac{L_{ion}}{n_0 (r_0)^2}
\label{xi}
\end{equation}
and the final radius of the wind $r_1$ is computed as a function of $N_H^{max}$:
\begin{equation}
N_H^{max}=\int_{r_0}^{r_1} n(r) dr
\label{nh}
\end{equation}

For each $m$-th slab the program analytically calculates all the input quantities needed to perform a relativistically-corrected photoionisation run. In the following, we first describe the procedure for $m=1$ and then we generalise it for $m>1$ slabs. 
WINE sets $N_H =\delta N_H^1$ and $L_{ion}, S_I, \xi_0, n_0, \alpha$ according to the values provided by the user.
The slab initial radius is $r_1^i=r_0$, while the final radius $r_1^f$ is as a function of $\delta N_H^1$ through Eq. \ref{nh}. We define a column density-averaged slab velocity $v_{avg}$ as $v(r_{avg})=v_0 \Big( \frac{r_0}{r_{avg}} \Big)^{\alpha_v}$, where $r_{avg}$ is the radius enclosing half of the slab column density: $\delta N_H /2 = \int_{r^i_1}^{r^{avg}_1} n(r) dr$. The procedure described in \cite{ltp20} is implemented to account for the special relativity effects in radiative transfer due to the bulk motion of the gas. Finally, $v_{turb}$ is defined as:
\begin{equation}
v_{turb} = max: \begin{cases} \delta v & \mbox{(velocity shearing)} \\ \sigma_{int} & \mbox{(intrinsic turbulence)} \end{cases}
\label{vturb}
\end{equation}
where $\delta v=|v(r_1^i)-v(r_1^{avg})|$ represents the variation of $v(r)$ within the slab, while $\sigma_{int}=0.1 \cdot v(r_{avg})$ and accounts for the intrinsic turbulence of the outflow. Given that here we focus on the study of the fastest wind components, we set a lower limit to $\sigma_{int}$ equal to 3000 $km s^{-1}$, in order to match the typical turbulence detected in the more distant (and perhaps less turbulent) Broad Line Region clouds \citep{kollatschny12}. Moreover, such lower limit is also consistent with what observed for UFOs (see e.g. \citealp{tcr11,grt13}. However $\sigma_{int}$ is a free parameter in WINE and can be changed by the user.
The \xstar\ run returns the (relativistic-corrected) transmitted spectrum from the first slab, corresponding to a column density $N_H=\delta N_H$ and, separately, the wind line emissivities (see below for more details).

To calculate the radiative transfer for the second slab, the transmitted spectrum from the first slab $S_{T,1}$ is given as input incident spectrum. Accordingly, the initial radius of the slab corresponds to the final radius of the previous slab, i.e. $r_2^i=r_1^f$, and the final radius $r_2^f$ is calculated in the same way of $r_1^f$. \xstar\ is run using $\delta N_H^2, n(r=r_2^i), S_{1-2},\xi(r=r_2^i)=\frac{L_{T,1}}{n(r=r_2^i) (r_2^i)^2}$, where $L_{T,1}$ is the luminosity of $S_{T,1}$. $v_{avg}$ and $v_{turb}$ are updated accordingly. This procedure is iterated over the $M$ slabs, up to $N_H^{max}$ and the absorption spectra up to $N_H^{max}$ are computed. 

\subsection{Wind emission}
\label{em}
In \xstar\ and in most of the current codes (see Sect. \ref{intro}) the gas emission is computed assuming null outflow velocity and a spherically symmetric geometry around the central luminosity source. In order to go beyond such assumptions, WINE models the emission profile via a Monte Carlo approach as a function of the wind kinematics and the geometry, adopting the same multilayer structure used to compute the transmitted spectrum.

For each slab, the code uses the \xstar-computed line emissivities $\zeta$, which are in units of $erg\ s^{-1}\ cm^{-3}$ and represent a "luminosity density" for a given transition within the slab.
The emission spectrum of the $m$-th slab is calculated according to the following steps:
\begin{itemize}
\item The volume of the slab $V_s$ is determined as a function of its geometry, i.e. $r^i_m, r^f_m, \theta_{out}, \theta_{in}$
\item The slab is populated with a high number $K$ of points, whose coordinates are randomly assigned with a Monte Carlo method inside $V_s$. \footnote{We find that $K=10^4$ allows to optimally sample the slab volume while still being computationally efficient.}
\item For each $k$-th point (where $k \in [1,K]$):
\begin{itemize}
    \item{An average volume $\delta V= V_s / K$ is assigned, and the luminosity $Z_i$ of the $i-th$ transition is calculated as $Z_i= \zeta_i \cdot \delta V$, where $i \in [0,I]$.}
\item{Using special relativity formulae, the line energy $E_i$ and luminosity $Z_i$ are projected along the LOS according to the so-called "relativistic beaming", which is described as \citep{ltp20}:
\begin{equation}
\begin{split}
& E_i^{LOS}=\psi \cdot E_i \\
& Z_i^{LOS}=\psi^3 \cdot Z_i
\end{split}
\label{en_proj}
\end{equation}
where $\psi=\frac{1}{\gamma (1-\beta cos(\theta))}, \gamma=\frac{1}{\sqrt{1-\beta^2}}, \beta=v_{avg}/c$, $v_{avg}$ represents the column density-averaged slab velocity and $\theta$ is the angle between the LOS and the coordinates of the $k$-th point.}
\end{itemize}
\item{The combination of $E_i^{LOS}, Z_i^{LOS}$ from all the lines (i.e. from $i=0$ to $=I$), gives the emitted spectrum of the $k$-th point.}
\item The sum of the spectra from $k=0$ to $K$ produces the slab emission spectrum.
\end{itemize}

Finally, the composition of the slab emission spectra for increasing $m$ gives the wind emission spectrum for increasing column density, up to $N_H^{max}$. To account for the self-absorption of such emission spectrum by the gas, at each $m$-th slab the radiation emitted by the inner slabs is absorbed according to the opacity $\tau$ of the present slab. Such opacity is computed as the ratio between the transmitted and the incident spectrum of that slab, accounting for relativistic effects as in Sect. \ref{abs}. Moreover, we introduce a volume filling factor $C_v$, so that only a fraction $0 \leq C_v \leq 1$ of the incoming emission is absorbed. We illustrate some examples in Sect. \ref{em_profiles}. The physical scenario of WINE is that of a single-phase gas column and, thus, $C_v=1$ would be expected. However, we allow for lower values to account for possible inhomogeneities, including clumpiness, which would allow part of the emitted spectrum to escape the wind unattenuated. In the highest signal-to-noise spectra, $C_v$ could be directly fitted through to the spectroscopic imprint of the self-absorption on the emission profiles (see Figs. \ref{vfig_xi3}, \ref{vfig_xi4}). More generally, hints on the degree of inhomogeneity of the gas can be obtained also indirectly, e.g. if the best-fit values of the wind absorption and emission are different, therefore suggesting different properties along the LOS with respect to the whole emitting region, or in case of temporary obscuring winds (such as those seen in NGC 5548 or MR 2251, \citealp{kaastra14,mao22}) due to denser clumps embedded into the intervening gas.

\begin{figure}
    \centering
    \includegraphics[width=0.9\columnwidth]{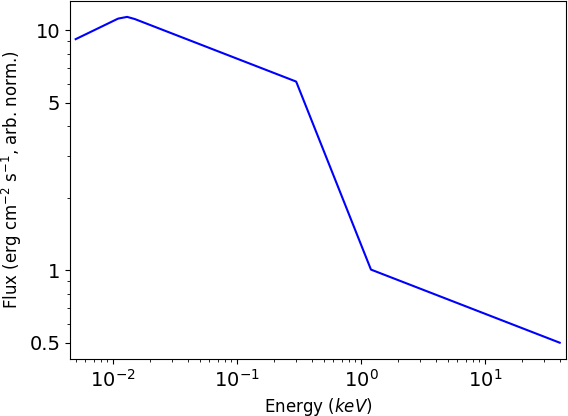}
    \caption{UV to X-ray SED of the NLSy1 I Zwicky 1.}
    \label{fig:izw_SED}
\end{figure}

\section{Wind spectra}
\label{outputs}
\subsection{Input SED and wind parameter space}

In this section we show some results obtained with WINE regarding the absorption and emission features emerging from an ionised gas with a range of physical properties. We consider the UV-to-X-ray SED of a well-studied Narrow-Line Seyfert 1 galaxy (NLSy1, \citealp{komossa08,tarchi11,rakshit17}), i.e. I Zwicky 1, which can be considered a representative highly-accreting AGN showing an UFO-related P-Cygni feature in its X-ray spectrum (\citealp{reeves19}; RB19 hereafter).
As reported in RB19 and shown in Fig \ref{fig:izw_SED} for clarity, the UV to X-ray spectrum is parameterised as a series of broken power laws with a peak in the far-ultraviolet regime and it is build from previous \textit{FUSE} \citep{fuse00,sahnow00} observations and the 2015 datasets from OM and EPIC instruments onboard XMM-\textit{Newton} \citep{jansen01,mason01,struder01,turner01}. It consists of a powerlaw with photon index $\Gamma=1.75$ below 12.5 eV and a second one with $\Gamma=2.2$ up to 300 eV. Between 300 eV and 1.2 keV $\Gamma=3.3$ to approximate the strong observed soft excess, while $\Gamma=2.2$ above 1.2 keV. The black hole mass and the 2-10 keV luminosity reported in RB19 are $M_{BH}=2.8 \cdot 10^7 M_{\odot}, L_{2-10}=5 \cdot 10^{43} erg\ s^{-1}$, implying $\lambda_{Edd} \approx 0.85$. For simplicity, we set a flat gas density profile (i.e. $\alpha=0$ in Eq. \ref{WINE:n_r}). We set the launching radius to a fiducial value of $r_0=50 r_G$ (see discussion in \citealp{lne21}) and $v_{turb}$ according to Eq. \ref{vturb}.
Here and in the following section we span the typical parameter space of X-ray UFOs, i.e. $N_H \in [10^{22}, 10^{24}], v_0 \in [0, 0.3] c, log(\xi_0) \in [3, 5]$. We consider a momentum-conserving wind and, consequently, we set the velocity index $\alpha_v=(2-\alpha)/2$. We set $v_{turb}$ to the (fiducial) lower limit of $3000 km/s$ to reduce the smearing and to better investigate the wind absorption and emission lines. Finally, we fix $\theta_{in}=0$ for simplicity.

According to these initial conditions, the wind is in a geometrically thin configuration and the scaling of $v, \xi$ with $r$ is limited, so it is possible to clearly appreciate the role of the different parameters. A useful quantity to characterise the wind geometrical thickness as a function of $N_H$ is $\Delta r = \frac{r(N_H)-r_0}{r_0}$, i.e., the difference between the radius enclosing a column density $N_H$ and the initial radius $r_0$, normalised by $r_0$. According to the density profile $\alpha$, $\Delta r$ can be expressed as:

\begin{equation}
\Delta r = \begin{cases} \frac{N_H}{n_0 r_{0,g}} \propto \frac{N_H}{\lambda_{Edd}} & \mbox{$\alpha=0$} \\ \exp{\frac{N_H}{n_0 r_{0,g}}} - 1 \propto \exp{\frac{N_H}{\lambda_{Edd}}} & \mbox{$\alpha=1$} \\ \frac{1}{( 1- \frac{N_H}{n_0 r_{0,g} (\alpha-1)} \big)^{\frac{1}{\alpha-1}}} -1 & \mbox{$\alpha \neq 1$} \end{cases}
\label{dr-cases}
\end{equation}

Where $r_{0,g} \equiv r_0/r_G$ is the launching radius in gravitational units. For $\alpha=0, \Delta r$ can be written as:
\begin{eqnarray}
\frac{N_H}{n_0 r_0} = \frac{N_H \xi_0 r_0}{L_{ion}} \approx 10^{-33} \frac{N_H \xi_0 r_{0,G}}{ \lambda_{ion}}
= 6 \cdot 10^{-3} \cdot \frac{N_H}{10^{24}} \frac{\xi_0}{10^5} \frac{r_{0,G}}{50}
\label{dr}
\end{eqnarray}
where we put $\lambda_{ion} \equiv L_{ion} / L_{Edd}=0.85$ as for I Zwicky 1. In this formula we omitted special relativity effects for simplicity (see \citealp{luminari23a} for more details).
The variation of ionisation and velocity along the wind column can be expressed as:
\begin{equation}
\xi(r)=\xi_0 (\Delta r + 1)^{\alpha-2}\\
\label{xi_r}
\end{equation}
\begin{equation}
v(r)=v_0 (\Delta r + 1)^{(\alpha-2)/2}\\
\label{v_r}
\end{equation}
so that $\xi(r) \approx \xi_0, v(r) \approx v_0$ in the whole parameter space probed in this paper, therefore confirming the negligible variation of the wind properties along the LOS.

\subsection{Absorption profiles}
\begin{figure}
\centering
\includegraphics[width=\columnwidth]{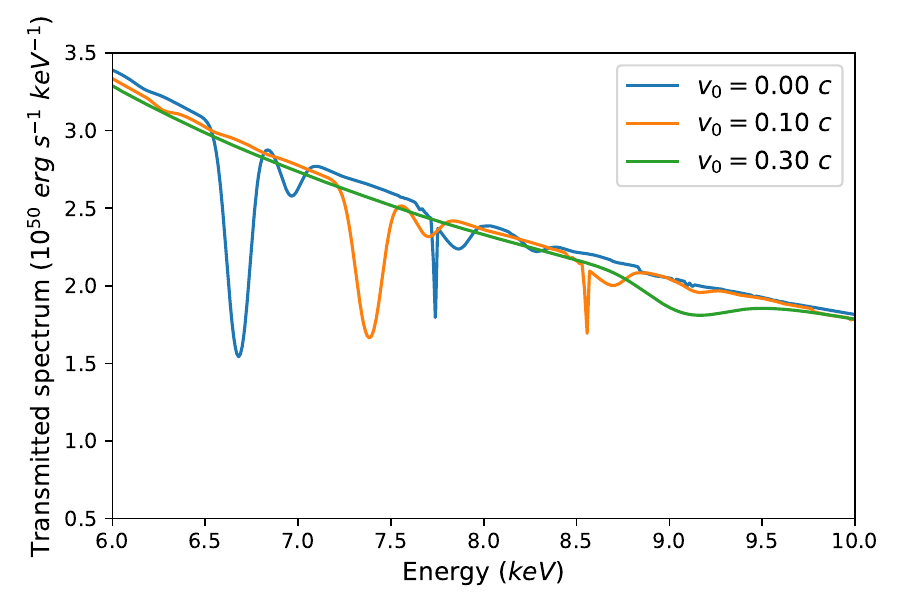}
\includegraphics[width=\columnwidth]{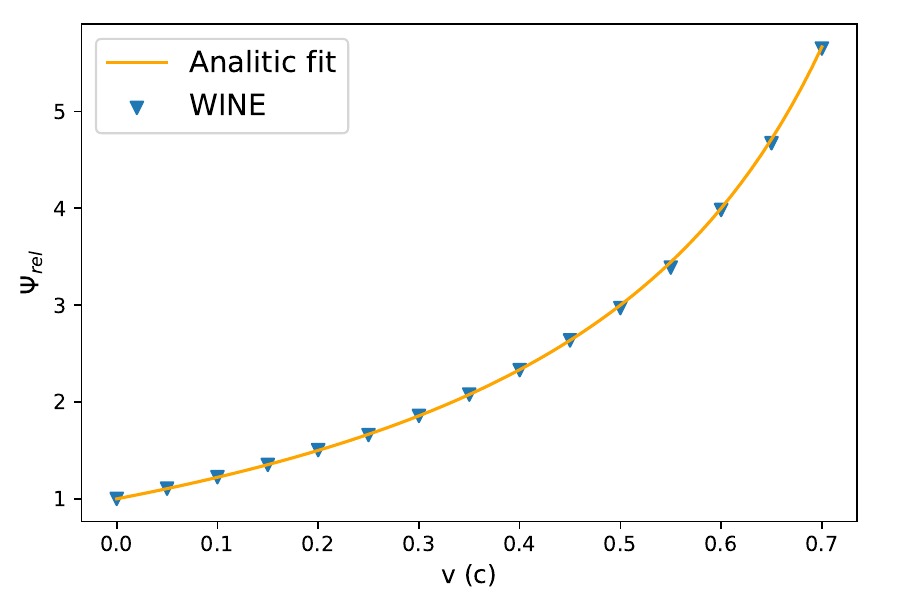}
\caption{Top: Absorption profiles for $v_0= 0.00, 0.10, 0.30 c$ (colour coding, see legend) and $N_H=10^{23}, log(\xi_0)=4$ in the rest-frame spectrum between 6-10 keV. For increasing $v_0$, absorption lines are increasingly blueshifted and show lower EWs, as expected from special relativity effects. Bottom: The ratio $\Psi_{rel}$ between intrinsic and observed gas opacity as a function of the gas velocity $v$ (in units of c).}
\label{4abs_Nh-v}
\end{figure}

\begin{figure*}[!ht]
\centering
\includegraphics[width=2\columnwidth]{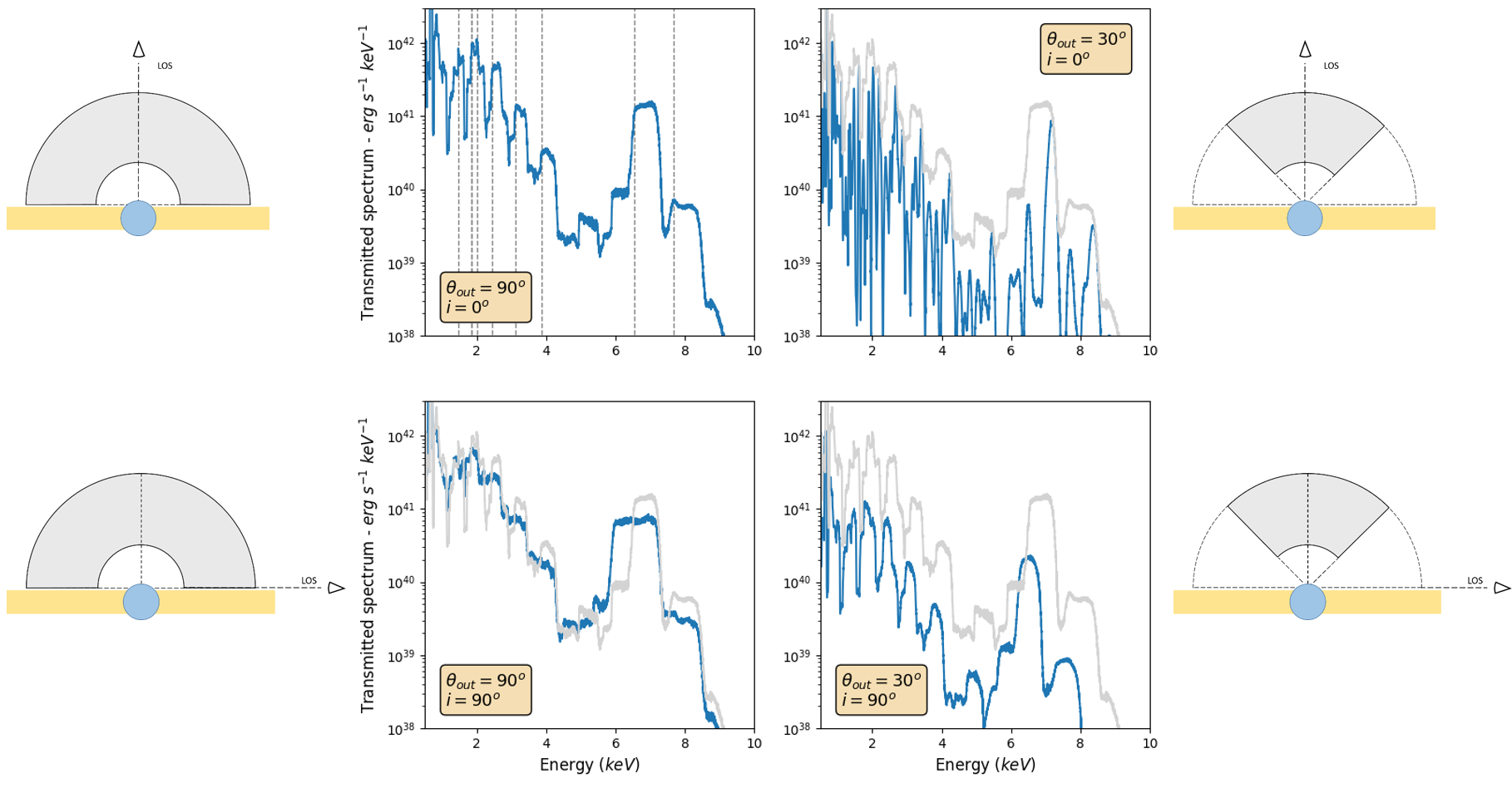}
\caption{Emission profiles for different wind geometries, assuming $log(\xi_0)=3, v_0=0.10 c, N_H=10^{23}$. Top row: wind seen "face on" ($i=0$) and $\theta_{out}=90, 30\ deg$ (left and right panel, respectively). Bottom row: "edge on" winds ($i=90\ deg$) and $\theta_{out}=90, 30$ (left and right panel, respectively). For comparison, the $i=0, \theta_{out}=90\ deg$ case is reported with light grey lines in all the panels.}
\label{geofig}
\end{figure*}

Fig. \ref{4abs_Nh-v}, top panel, shows the absorption spectrum for $v_0=0.05, 0.10, 0.30 c$ and $N_H=10^{23}, log(\xi_0)=4$. The strongest absorption profiles are due to the Ly$\alpha$, Ly$\beta$ lines of Fe XXIV and XXV, which are the most abundant Fe ions at such ionisation \citep{kallman04}. Due to special relativity effects, for increasing outflow velocities absorption lines are both blueshifted and have lower equivalent widths, as discussed in \cite{ltp20}. Moreover, the turbulent velocity $v_{turb}$ increases for increasing $v_0$ (following Eq. \ref{vturb}) and strongly modifies the absorption profiles, blending together the Ly$\alpha$ lines of Fe XXV and XXVI for $v_0=0.30\ c$ at energies around 9 keV.

Due to the relativistic effects, when inferring the column density of the gas from its measured opacity in observed spectra it is necessary to include a correction to get the intrinsic gas opacity and, then, the intrinsic $N_H$. Fig. \ref{4abs_Nh-v}, bottom panel, shows the ratio $\Psi_{rel}$ between intrinsic and observed opacity as a function of the gas velocity $v$ (in units of c) computed with WINE (blue diamonds), assuming $v$ parallel to $r$ for simplicity. This relation is found to be well described by the following equation: 
\begin{equation}
\Psi_{rel}=\frac{1+v/c}{1-v/c}
\label{eq:abs-rel}
\end{equation}
which can be used to derive the intrinsic $N_H$ from the $N_H$ obtained from the measured (i.e. apparent) opacity.

\begin{figure*}
\centering
\includegraphics[width=1.9\columnwidth]{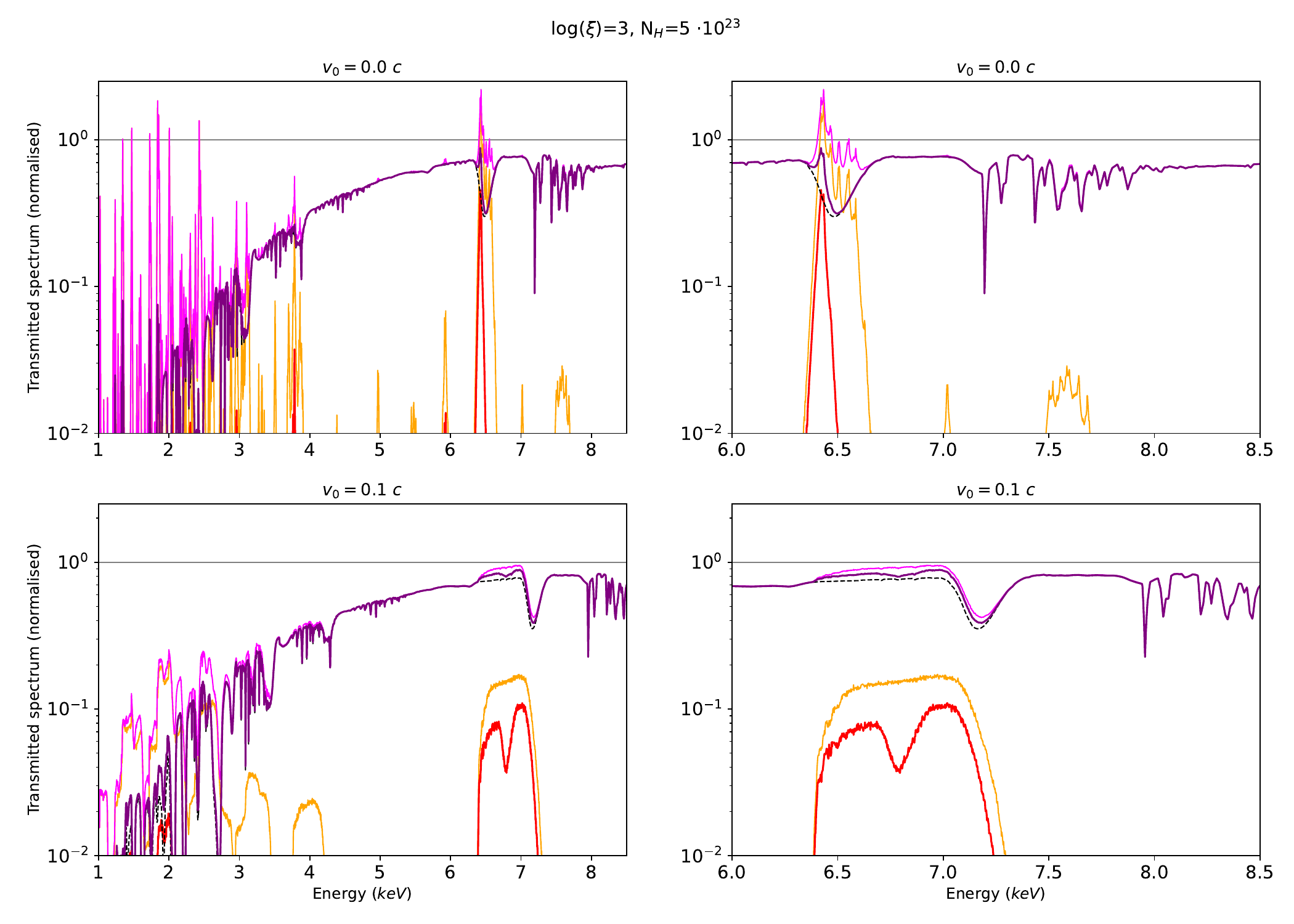}
\caption{WINE spectrum for $log(\xi_0)=3, N_H=5 \cdot 10^{23}$ and $v_0=0.0, 0.1\ c$ (top and bottom row, respectively). Left panels show the E=1 - 8.5 keV spectrum, while right panels are zoom-in above E=6 keV. Orange(red) lines represent the emission spectrum and pink(purple) the total (absorption + emission) spectrum for $C_v=0$(1). The absorption spectrum is shown with black dashed lines. For ease of visualisation, the spectra are divided by the incident continuum.}
\label{vfig_xi3}
\end{figure*}

\begin{figure}
\centering
\includegraphics[width=0.9\columnwidth]{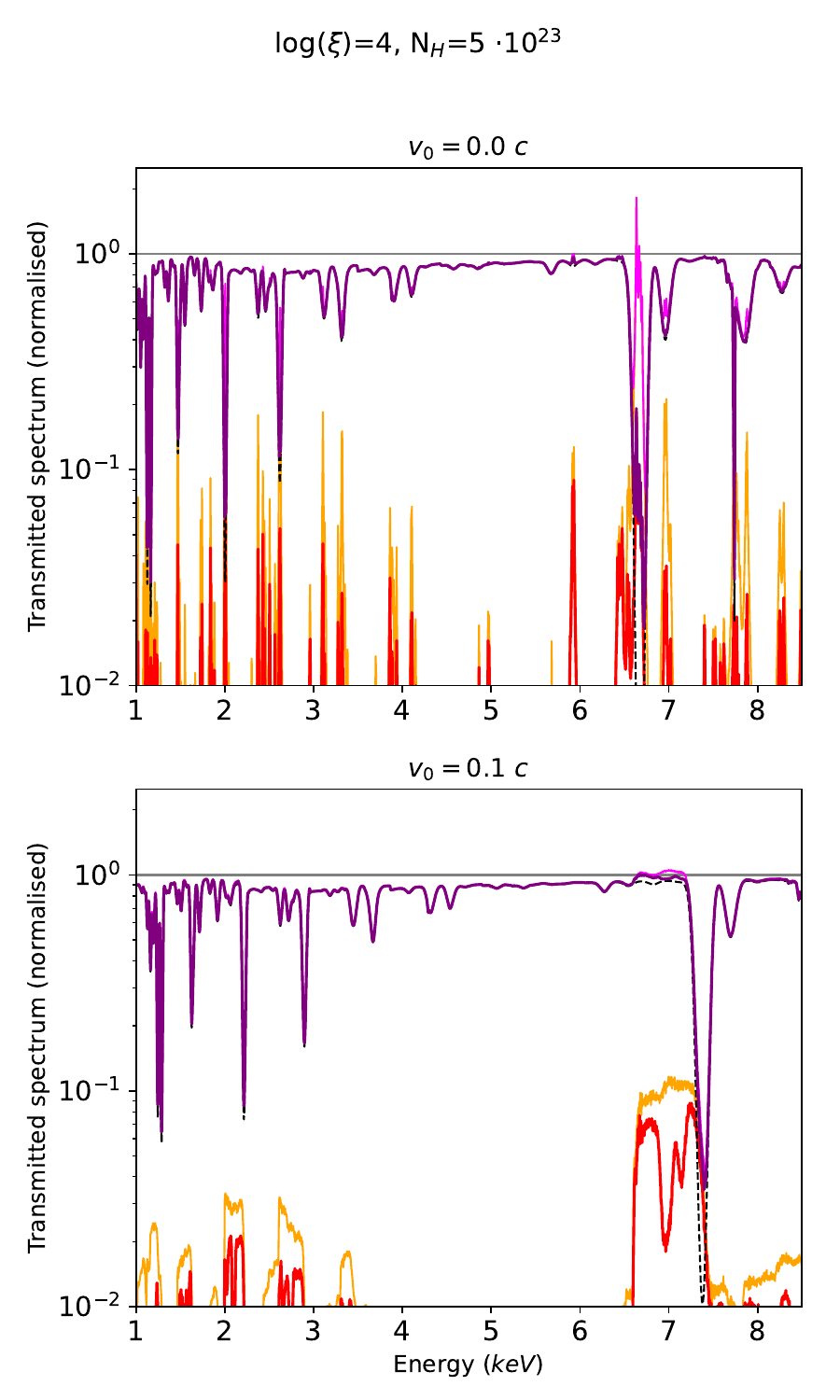}
\caption{WINE spectrum for $log(\xi_0)=4, N_H=5 \cdot 10^{23}$ and $v_0=0.0, 0.1\ c$ (top and bottom panel, respectively). Colour coding as in Fig. \ref{vfig_xi3}.}
\label{vfig_xi4}
\end{figure}

\subsection{Emission profiles}
\label{em_profiles}

Fig. \ref{geofig} shows the emission profiles for $log(\xi_0)=3, v_0=0.1 c, N_H=10^{23}$ for different wind geometries. According to Eq. \ref{en_proj} the projected velocity $v_{proj}$ can be expressed as: $v_{proj} = \frac{v_{out}}{\gamma (1- \beta \cos \theta)}$, where $\theta$ is the angle between the LOS and the gas velocity (which depends on $\theta_{out}, \theta_{in},i$, see Fig. \ref{4cone}). When $\theta=0\ deg$, i.e. the gas is directed radially outward and $v_{out}$ is parallel to the LOS, $v_{proj}$ is the highest, $v_{proj}= \frac{v_{out}}{\gamma (1- \beta)}$, while when the velocity is opposed to the LOS, i.e. $\theta=180\ deg$, $v_{proj}$ is the lowest, $v_{proj}= \frac{v_{out}}{\gamma (1+ \beta)}$.

In the top left panel we show the case for $\theta_{out}=90\ deg, i=0$: the transmitted flux is maximised, since the emitting volume is the largest possible, as shown in the cartoon (i.e., a hemisphere filling the solid angle above the accretion disc) and the LOS coincides with the symmetry axis, resulting in a beamed emission towards the observer (see Eq. \ref{en_proj}). The accretion disc is thus seen "face on" and $\theta$ ranges from 0 to 90 $\deg$. 

The emission profile is the combination of the lines due to several ions, with the strongest ones originated from the transitions with the highest oscillator strengths for the most abundant ions. We highlight with vertical dashed lines the rest-frame wavelength of some of them; from left to right, they are due to Mg XXII, Si XIII, Si XIII, Si XIV, S XV, Ar XVII, Ca XIX, Fe XXII Ni XXIV. The emission lines extend blue-ward of their rest-frame energies, spanning a range of projected velocity from $\frac{v_{out}}{\gamma}$ (for the gas at $\theta=90 \deg$) to $\frac{v_{out}}{\gamma(1+ \beta)}$ ($\theta=0 \deg$).

In the top right panel of Fig. \ref{geofig} we set the cone opening angle $\theta_{out}=30\ deg$, and we keep the inclination $i=0$. The emitted luminosity is lower than that found in the previous case (plotted again with grey lines for comparison) and the low-energy (low-velocity) side of each emission peak is reduced, since there is no gas at $30 \deg < \theta_{out} < 90\ deg$.
In the $\theta_{out}=90, i=90\ deg$ case (bottom left panel of Fig. \ref{geofig}), the accretion disc is instead seen "edge on", and then $\theta$ ranges from 0 to $180\ deg$. As a consequence, the spectral features are broadened over a larger energy interval than in the "face on" case, since they are projected over an increased velocity space. Finally, for $\theta_{out}=30, i=90\ deg$ (bottom right panel) the emission region is outside the LOS and $\theta$ ranges from 60 to 120 $deg$. This explains the lower $v_{proj}$ and the overall decrease of the transmitted flux, since the relativistic beaming reduces the radiation projected along the LOS.

Wind emission spectra for $log(\xi_0)=3$ and 4 are shown in Fig. \ref{vfig_xi3} and \ref{vfig_xi4}, respectively. In both Figures, top and bottom panels are for $v_0=0.0$ and 0.1 c, while in all cases $\theta_{out}=90, i=0\ deg, N_H= 10^{24}$. Orange(red) lines indicate the emission profiles for $C_v=0$(=1) and pink(purple) lines the composition of such spectrum with the absorption one. For reference, absorption spectra are reported with black dashed lines. For increasing $v_0$, the gas cross section decreases due to relativistic effects and emission features are spread over a larger observed energy interval. As a consequence of both these effects, the emitted flux per unit energy is lower for increasing $v_0$. As we will show in Sect. \ref{pcyg}, this will have important implications on the detectability of emission features in the total (i.e., emission+absorption) wind spectrum. By comparing Fig. \ref{vfig_xi3} and \ref{vfig_xi4}, it can be seen that the higher gas ionisation directly translates into lower gas opacity. For $log(\xi_0)=4$, strongest transitions in the hard X-ray band (E>5 keV) are those of Iron XXV K$\alpha$ and K$\beta$  (E=6.7, 7.8 keV, respectively) and Iron XXVI K$\alpha$ (E=6.9 keV).
The attenuation of the emitted radiation by the gas self-absorption is mostly effective both for high gas opacity and for low velocity. The latter effect is due to the fact that absorption and emission lines are closer in the velocity space and, thus, the weakening of the emerging emission is maximised.  
The narrow absorption features at rest frame energies E=7-8 keV and, for $log(\xi)=3$, also between E=3-5 keV, are due to resonances of the photoelectric cross section of several abundant elements, most notably Iron \citep{kallman04}, which are not convolved for the gas turbulent velocity in XSTAR (T. Kallman, priv. comm.). The high gas opacity for $log(\xi)=3, N_H=5 \cdot 10^{23}$ leads to strong absorption and emission in the soft band (E< 5keV), virtually leading to a complete absorption of the incident continuum.

\section{Wind emission detectability}
\label{pcyg}

\begin{figure}
\centering
\includegraphics[width=0.9\columnwidth]{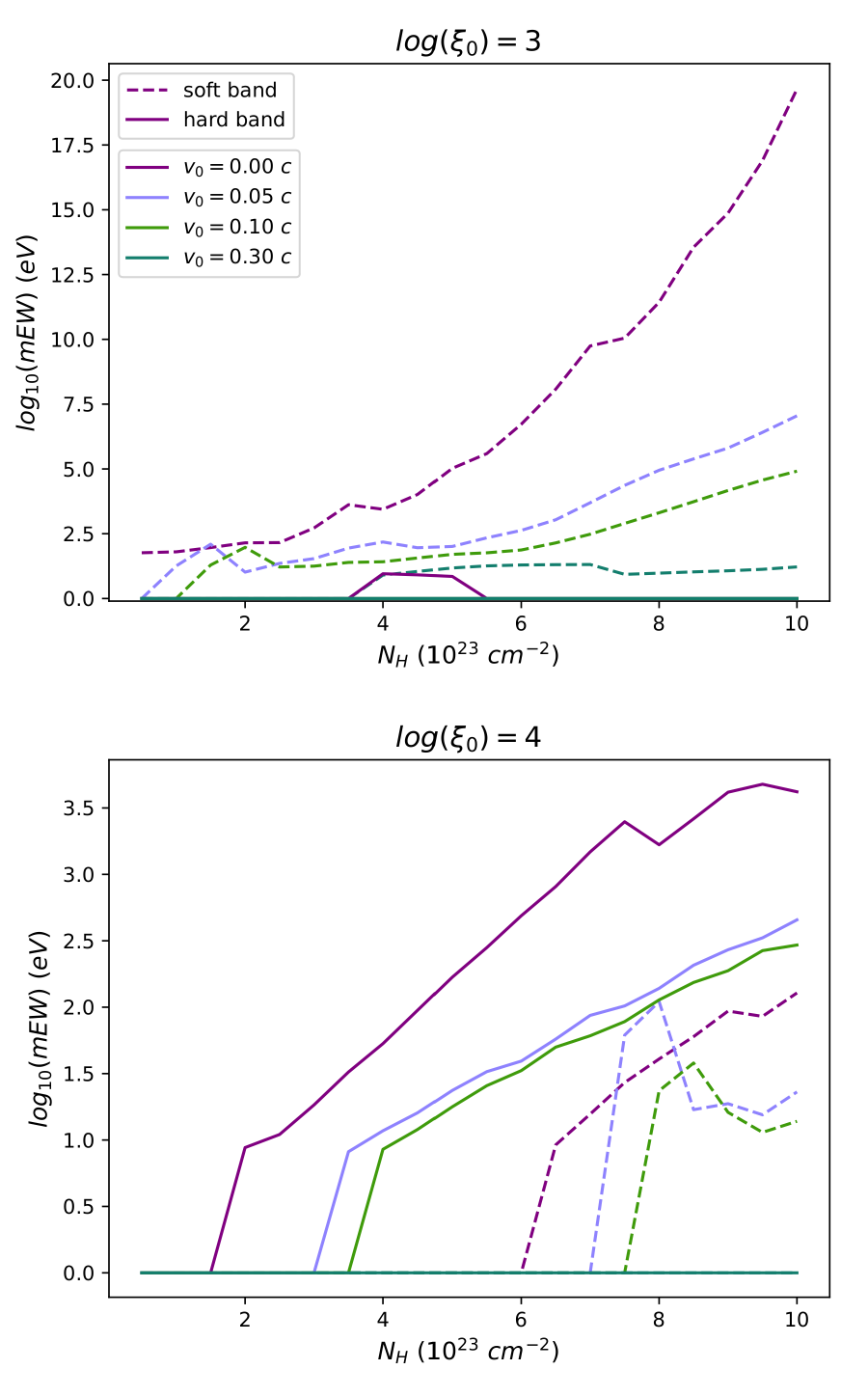}
\caption{Logarithm of the Median Equivalent Width (mEW) of the wind emission as a function of $N_H$, plotted for increasing $v_0$ (colour coding, see legend). Soft and hard X-ray band, corresponding to 0.5 - 2.5 and 5-15 keV, respectively, are shown with dashed and solid lines. Top and bottom panels correspond to $log(\xi_0)=3,4$ respectively.}
\label{4elmtable_43}
\end{figure}

We explore the disc wind parameter space to evaluate the strength of the emission features in the wind spectrum. Starting from the input parameters presented in Sect. \ref{outputs}, we simulate with WINE a range of column density $N_H \in [5,20] \cdot 10^{23}$ for $log(\xi_0) \geq 3$, in order to optimally sample the properties of the P-Cygni profiles observed in UFOs (see Sect. \ref{intro}). We set $\theta_{out}=90\ deg, \theta_{in}=0, i=0 \deg$ in order to maximise the wind emission, and $C_V=1$ as suggested by such high-covering factor geometry. We set an energy resolution of 5 $eV$ as a fiducial reference value for the {\it Resolve} microcalorimeter onboard {\it XRISM}.

We define the emission {\it detectable} if the flux of the emission spectrum is higher than the absorption one in at least two adjacent energy bins. We select two energy bands of interest, a soft band from 0.5 to 2.5 keV and a hard band between 5 and 10 keV. 
Fig. \ref{4elmtable_43} shows the median Equivalent Width (mEW) of the emission features, derived by the following procedure. We scan the soft and hard energy bands looking for intervals of two or more bins satisfying the detectability requirement. For each interval, we compute the EW of the emitted spectrum with respect to the transmitted one. Finally, we calculate the median value of the distribution of EWs for all the intervals. 
We focus on mEW, rather than the total EW, since we are primarily interested on the detectability of such features. Although an emission spectrum whose flux is slightly higher than the absorption one in a large energy range will have a non-negligible EW, it would be difficult to detect due to the low contrast with respect to the underlying absorbed continuum spectrum. Conversely, emission features which are much stronger than the absorbed continuum, even if concentrated in few energy bins, will lead to an easily detectable spectral feature. Finally, null mEW values indicate that the detectability condition has not been matched in the whole energy band.
The mEW are reported as a function of $N_H$ for both the soft and the hard bands (dashed and solid lines, respectively). For $log(\xi_0)=3$ (top panel) emission is detectable in the soft band in winds with $v_0$ up to 0.3 $c$. As expected, the emission features are stronger for lower velocities, since the relativistic effects are lower (i.e. the overall interaction rate between radiation and gas is higher) and the emission profiles are neither smeared nor blueshited. In the hard band, the detectability threshold is not met for any $v_0$, except for $v_0=0$ and $4 \cdot 10^{23} \leq N_H \leq 5 \cdot 10^{23}$. This is due to two Fe XIX and Fe XXIII emission lines at E=6.4 keV which, however, get re-absorbed for higher gas column densities.
Finally, for $log(\xi_0)=4$ (bottom panel), most of the opacity of the highly-ionised gas falls in the hard band. Therefore, emission is more detectable in such band than in the soft one. For $v_0=0.3 c$ the detectability requirement is never met, due to strong suppression of the emission profiles by the relativistic effects, as discussed above. Finally, we do not plot the results for $\xi \geq 5$, since for such high ionisation the detectability threshold is never met.

\section{\textit{XRISM Resolve} simulations}
\label{resolve-sim}

\begin{figure*}
\centering
\includegraphics[width=1.85\columnwidth]{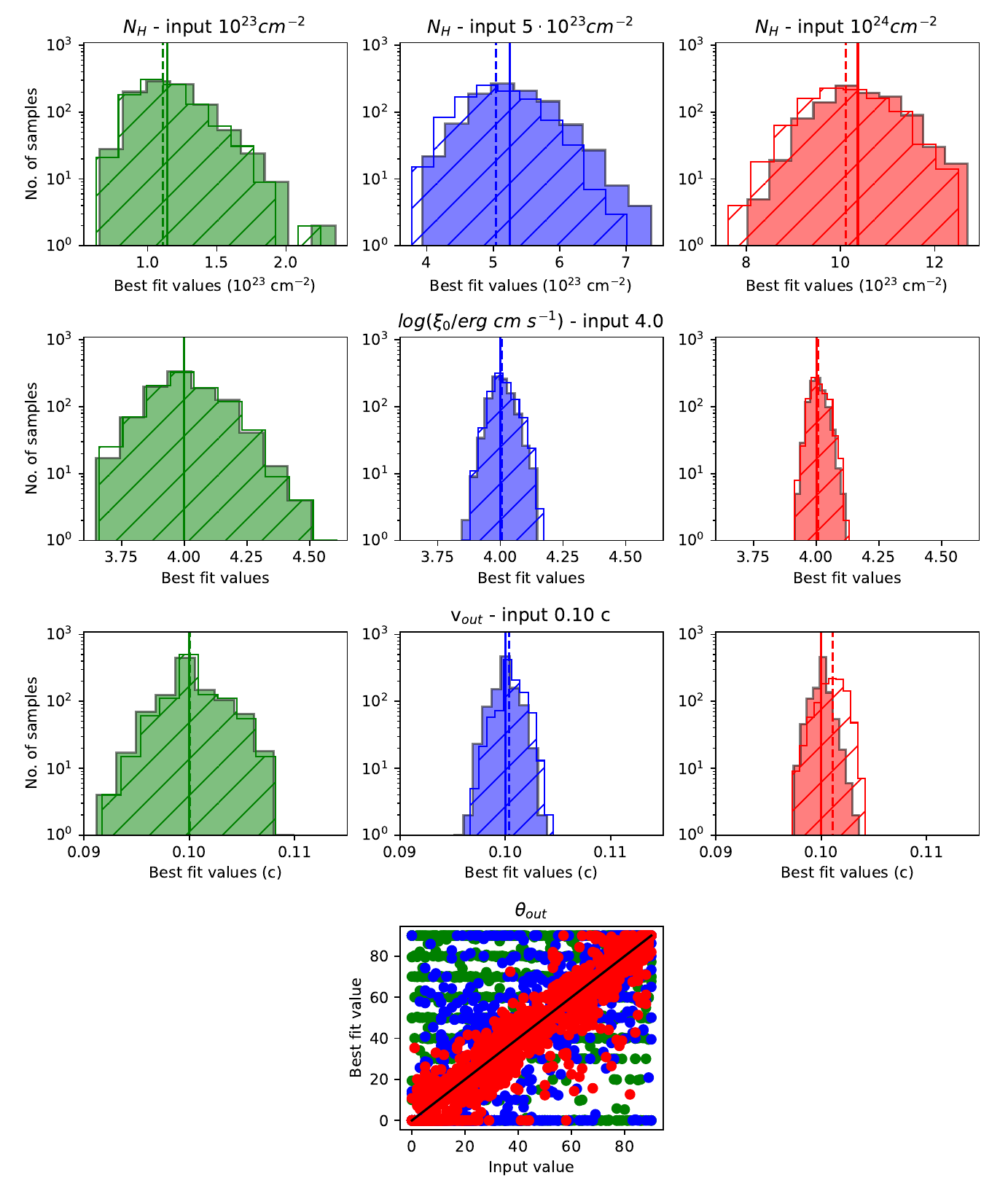}
\caption{Distribution of the best fit values of 1000 {\it XRISM Resolve} simulations, using as input values $log(\xi)=4.0, v_0=0.10 c$ and three different $N_H=10^{23}, 5 \cdot 10^{23}, 10^{24}$ (left, centre and right column, respectively). Top, centre and bottom row report the best-fit values for $N_H, log(\xi), v_0$, respectively. Hatched and filled histograms correspond to the distributions before and after the inclusion of the emission component, respectively. Dashed and solid lines are the median values of the two distributions. Finally, last panel reports the best-fit values of $\theta_{out}$ as a function of the input ones.}
\label{izw-3.5}
\end{figure*}

\begin{figure*}
\centering
\includegraphics[width=1.8\columnwidth]{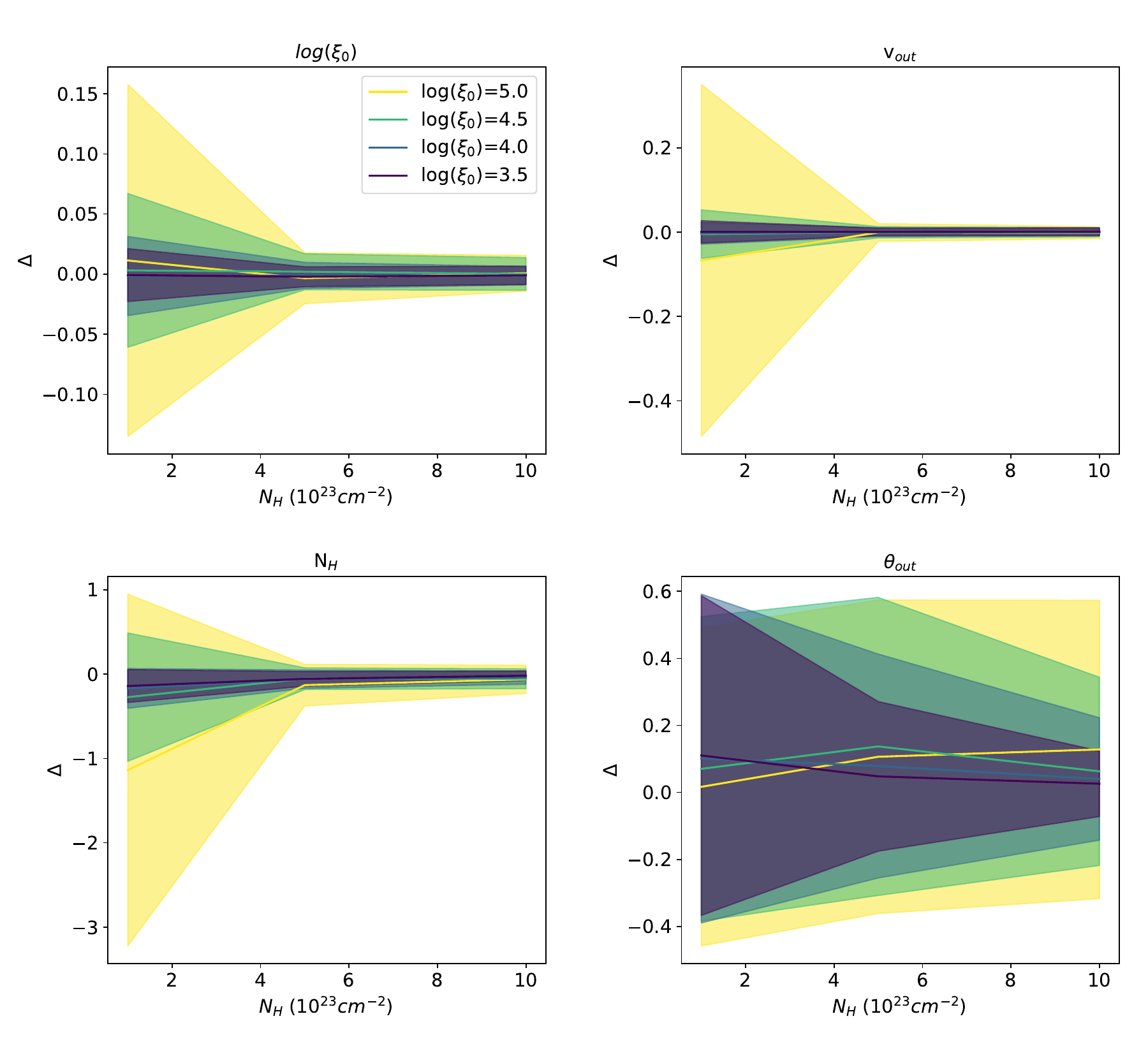}
\caption{Ratio $\Delta$ (rescaled to 0; see Eq. \ref{eq_sigma}) for \textit{XRISM Resolve} simulations between input and best fit values, as a function of the input $N_H$. The shaded areas show the corresponding standard deviation. The colour coding corresponds to the different input $log(\xi_0)$ (see legend). From left to right and top to bottom, the different panels correspond to the best fit distributions of $log(\xi), v_0, N_H, \theta_{out}$, respectively. For ease of visualisation, we only report the results after the inclusion of the emission component. Note that, in the last panel, only those simulations with input $\theta_{out}>45 \deg$ are included.}
\label{fig_simul}
\end{figure*}

\subsection{NLSy1 SED}
\label{izw-sim}
In this Section we probe the capability of the \textit{Resolve} microcalorimeter onboard \textit{XRISM} to constrain the wind emission properties. To do so, we build a table of WINE spectral templates that is provided as input to the \textit{xspec} fitting software \citep{xspec} as table models.
We focus on the following UFO parameter space:
\begin{itemize}
\item $log(\xi_0) \in [3.0,5.5]$, 0.25 step
\item $v_0 \in [0.00,0.40] c$, 0.025 step
\item $N_H \in [0,2 \cdot 10^{24}]$, $10^{23}$ step
\item $\theta_{out} \in [0,90 deg]$, $10 deg$ step
\item $i \in [0,90 deg]$, $10 deg$ step
\end{itemize}
and we set all the other properties as in Sect. \ref{outputs}
We note that $\theta_{out}=0$ implies a null volume of the wind cone (see Fig. \ref{4cone}) and therefore corresponds to a null emission spectrum. Since for given $\xi_0, N_H$, the flux of the emitted radiation is proportional to $C_f$, the strongest emission spectra will be associated with wide-angle winds. To better approximate the attenuation of the emitted radiation in these cases, we set a constant $C_v=1$ throughout the parameter space. 

Through \textit{xspec} we simulate \textit{Resolve} spectra with the following input model:
\begin{equation}
WINE_{abs} \cdot powerlaw + WINE_{em}
\end{equation}
where $WINE_{abs}$ and $WINE_{em}$ represent the WINE absorption and emission spectrum, respectively. Note that the gas self-absorption is accounted for during the creation of the emission spectra and, thus, it is already implemented in the $WINE_{em}$ component. The powerlaw normalisation is tuned to give a 2-10 flux of $10^{-11} erg\ cm^{-2}\ s^{-1}$, representative of the bright targets of the performance verification (PV) phase of \textit{XRISM} \footnote{The list can be found at: \url{https://xrism.isas.jaxa.jp/research/proposer/approved/pv/index.html}}.

Using this model, we simulate sets of 1000 spectra of 100 ks with several input values for both the ionisation, $log(\xi)=3.5$, 4.0,4.5,5.0, and the column density, $N_H (10^{23}$)=1.0, 2.5, 5.0, 7.5, 10.0, and with $v_0=0.1 c$. For each observation, we randomly assign a value for $\theta_{out}$, the opening angle of the cone, while for simplicity we fix $i=0, \theta_{in}=0$.
The spectra are fitted using a two-step approach:
\begin{itemize}
\item First, we fit the spectrum only with an absorption component ($WINE_{abs} \cdot powerlaw$), accurately scanning the parameter space by computing the error and the contour plots associated to each free parameter, i.e. $log(\xi_0), N_H, v_0$.
\item Once a best-fit model is obtained, we include an emission component where the values of $log(\xi_0), N_H, v_0$ are tied to those of the absorption component (since emission and absorption are generated by the same medium), and with $\theta_{out}$ free to vary. Then, we fit leaving all the parameter free to vary, and we scan the parameter space as above.
\end{itemize} 

We use up-to-date response and background files for point source observations for a closed Gate Valve configuration, available through the {\it XRISM} mission website\footnote{\textit{Resolve} files can be found at \url{https://xrism.isas.jaxa.jp/research/proposer/obsplan/response/index.html}}. 
Simulated spectra are binned to a minimum of 50 counts per bin and $\chi^2$ statistic is adopted. The energy range between 2.0 and 10.0 keV is considered. We verified that the results are the same if a binning with a fixed energy width of 10 eV and C-statistic \citep{cash76} are adopted.

Fig. \ref{izw-3.5} shows the distributions of the 1000 best fit solutions for input $log(\xi)=4$. For simplicity, we report only the results for input $N_H=10^{23}, 5 \cdot 10^{23}, 10^{24}$ (left, centre and right columns, respectively). Top, centre and bottom panels report the best fit values of $N_H, \xi, v_0$, respectively. Hatched and filled histograms represent the distributions before and after the inclusion of the emission component in the fitting model. The last panel shows the best fit values for $\theta_{out}$ as a function of their input values. The black diagonal line shows a 1:1 correspondence.
Increasing the input $N_H$ (left to right) leads to narrower best-fit distributions, since the wind opacity is higher and, thus, tighter constraints can be achieved. The values before and after the inclusion of the emission component basically overlap for $N_H=10^{23}$, suggesting that the emission component is subdominant in these fits. This is confirmed by the distribution of the values of $\theta_{out}$, which is spread over a broad range of the parameter space. Note that the clustering of the output values at steps of 10 $\deg$ is simply due to the resolution of the table model.
The impact of the emission component grows with increasing input $N_H$, as demonstrated by the increasing distance between the hatched and filled histograms and the tighter correspondence between input and output $\theta_{out}$. To provide a more quantitative evaluation of the ability in recovering the input parameters, we compute the average $\Delta$ as:
\begin{equation}
\Delta= 1- \langle \frac{V_{fit}}{V_{in}} \rangle
\label{eq_sigma}
\end{equation}
i.e. the ratio between the best-fit and the input values, $V_{fit}, V_{in}$ respectively, rescaled to 0. Figure \ref{fig_simul} shows $\Delta$ for input $log(\xi)=3.5,4.0,4.5,5.0$ as a function of the input $N_H$. The spread corresponds to the standard deviation of the distributions of the best fit values. For plotting purposes, we only report the values after the inclusion of the emission component. As expected, $\Delta$ is smaller for increasing $N_H$ and decreasing $\xi$, i.e. when the opacity of the wind increases and, thus, the fitting results are more constrained. Since we are mainly interested in the characterisation of wide angle outflows, which are the most likely ones in producing P-Cygni profiles, in the computation of $\Delta$ for \tout\ we include only the simulations with input $\theta_{out} \geq 45 \deg$ ($\approx 50\%$ of the total). See Appendix \ref{xrism-sims-appendix} for the computations with the full data set.

\begin{figure}
\centering
\includegraphics[width=\columnwidth]{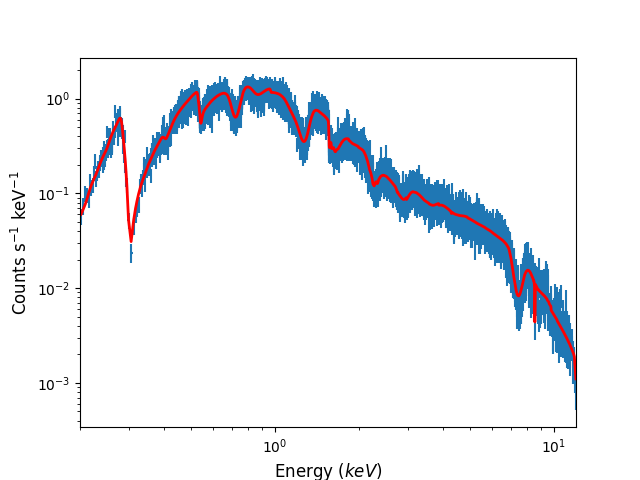}
\caption{Simulated 100ks \textit{XRISM Resolve} spectrum (blue points) for a wind with $log(\xi)=4, v_0=0.10 c, N_H=5 \cdot 10^{23}, \theta_{out}=72 \deg$ and with the Gate Valve open. The best fit model is indicated as a red line. Spectrum is rebinned for plotting purposes.}
\label{fig:resolve-spec}
\end{figure}

To give an idea of the capabilities of \textit{Resolve} with the Gate Valve open, Fig. \ref{fig:resolve-spec} shows a simulated 100ks \textit{Resolve} spectrum with input parameters $log(\xi)=4, v_0=0.10 c, N_H=5 \cdot 10^{23}, \theta_{out}=72 \deg$ (blue points), together with the fitted model (red line), with best fit values $log(\xi)=4.008 \pm 0.006, v_0=0.1004 \pm 0.001 c, N_H=4.9^{+0.4}_{-0.1} \cdot 10^{23}, \theta_{out}=71^{+11}_{-16} \deg$.

\subsection{Powerlaw SED}
\label{pow-sim}
In this Section, we adopt a power-law function as input SED and run new simulations to assess its impact on the simulated spectra. We do so to better compare our results with several X-ray observations reported in the literature, where UFOs have been analysed through photoionisation models assuming such an SED (see e.g. \citealp{tcr10, grm15,reeves23}).

We set as reference values a 2-10 keV luminosity of $L_{2-10}=10^{43} erg s^{-1}$ and a black hole mass $M_{BH}=10^7 M_{\odot}$, thus implying a value $\lambda_{Edd} \equiv \frac{L_{bol}}{L_{Edd}} \approx 0.1$ for an X-ray bolometric correction of $\approx 30$ \citep{lusso12}, a typical value for the bulk of AGN population in the local Universe ($z \lesssim 0.1$, see e.g. \citealp{veron-cetty06}). We set a power-law spectral photon index $\Gamma=1.8$, again as representative of the local population (see e.g. \citealp{pico05,tcr11} and references therein). As above, we assume an initial radius $r_0=50 r_G$ and a flat density profile ($\alpha=0$). The parameter space and the fitting strategy are as above. 

\begin{figure}
\centering
\includegraphics[width=\columnwidth]{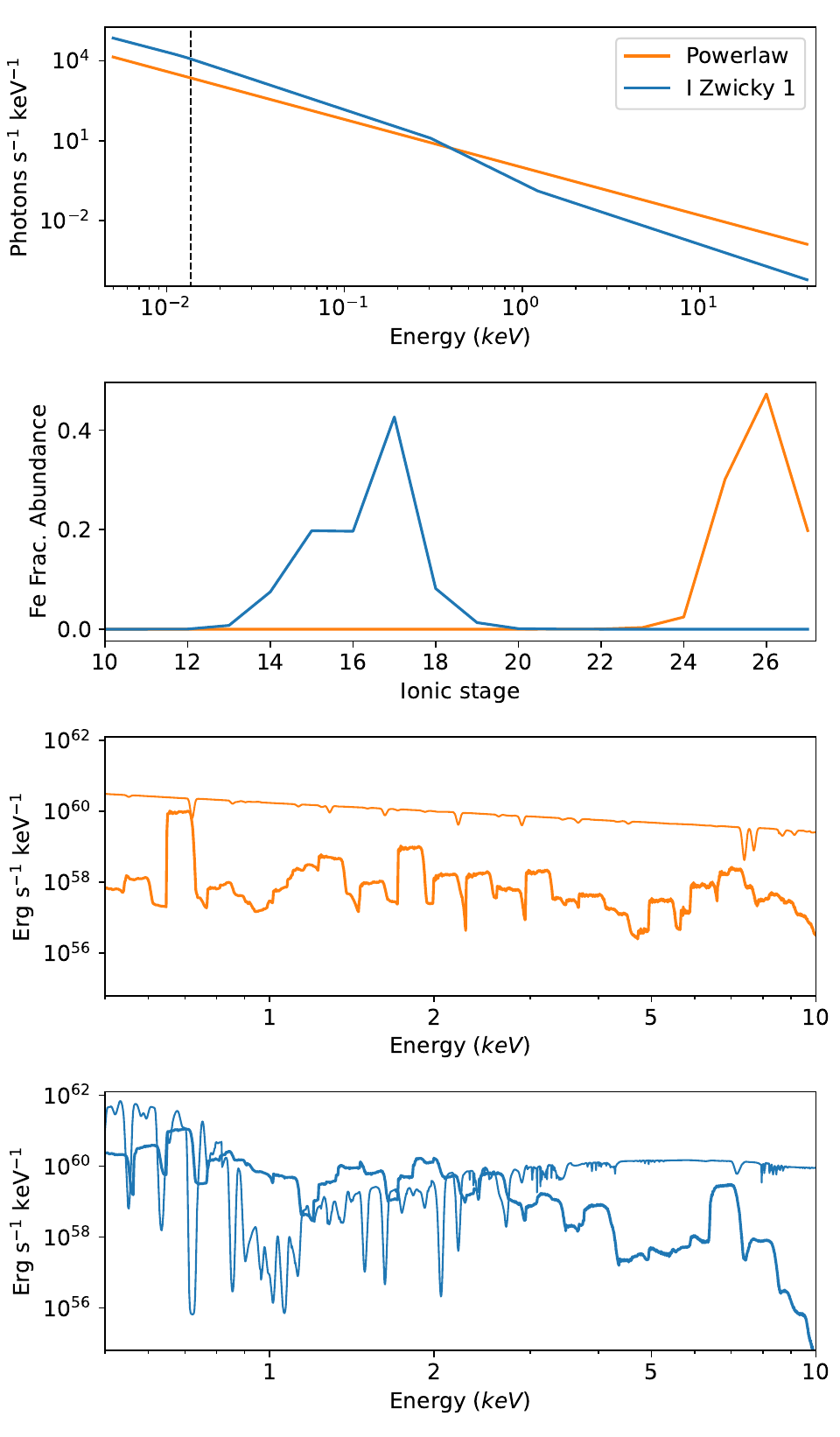}
\caption{First panel: comparison between the powerlaw SED (orange) and the NLSy1 one (blue), in units of Photons s$^{-1}$ keV$^{-1}$ (arbitrary normalisation). Second panel: Iron ionic fractional distribution for $log(\xi)=3, N_H=5 \cdot 10^{23}, v_0=0.1 c$ and the two different SED (colour coding as above). Third and fourth panels: absorption and emission spectra (thin and thick lines, respectively; units of Erg s$^{-1}$ keV$^{-1}$) assuming the powerlaw and the NLSy1 SED, respectively, and same $log(\xi), N_H, v_0$ as in the second panel.}
\label{sedcomparison}
\end{figure}

We compare this SED and the NLSy1 one in Fig. \ref{sedcomparison}, top panel, in units of Photons s$^{-1}$ keV$^{-1}$. Their relative ratio is set as to give the same $L_{ion}$. The black vertical line corresponds to 13.6 eV, i.e. the lower bound of $L_{ion}$. As expected, for a given $L_{ion}$, the NLSy1 SED shows more photons below 0.3 keV. For given $\xi$, this leads to a higher ionised gas for the powerlaw SED, as can be seen in the second panel, where we compare the distributions of the Iron ionic fractional abundances for $\xi=10^3, N_H=5 \cdot 10^{23}, v_0=0.1 c$ using the two SEDs. While with the NLSy1 SED the most abundant ion is Fe XVII, with the powerlaw SED it is Fe XXVI, and 20\% of iron is totally ionised (Fe XXVII). This can be also seen in the spectra of Fig. \ref{sedcomparison}, shown for both cases. We refer to \cite{nicastro99b} for further discussion on the difference between the NLSy1-like and powerlaw SEDs for the gas ionisation. 

We simulate sets of 1000 observations with \textit{Resolve}, following the same procedure of above. Unsurprisingly, we obtain that the gas emission is totally unconstrained for $log(\xi) \geq 4$, being the emission much weaker.

\begin{figure}
\centering
\includegraphics[width=\columnwidth]{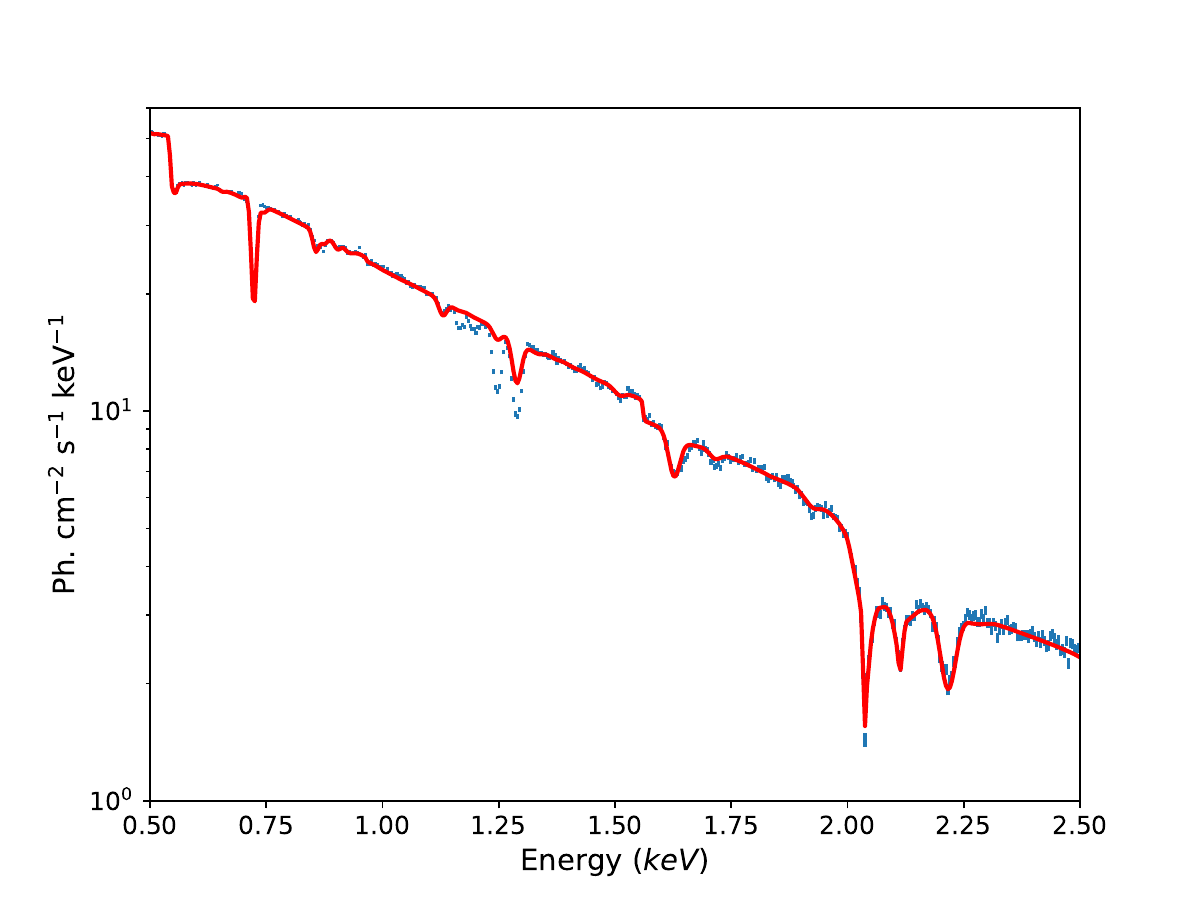} \\
\includegraphics[width=\columnwidth]{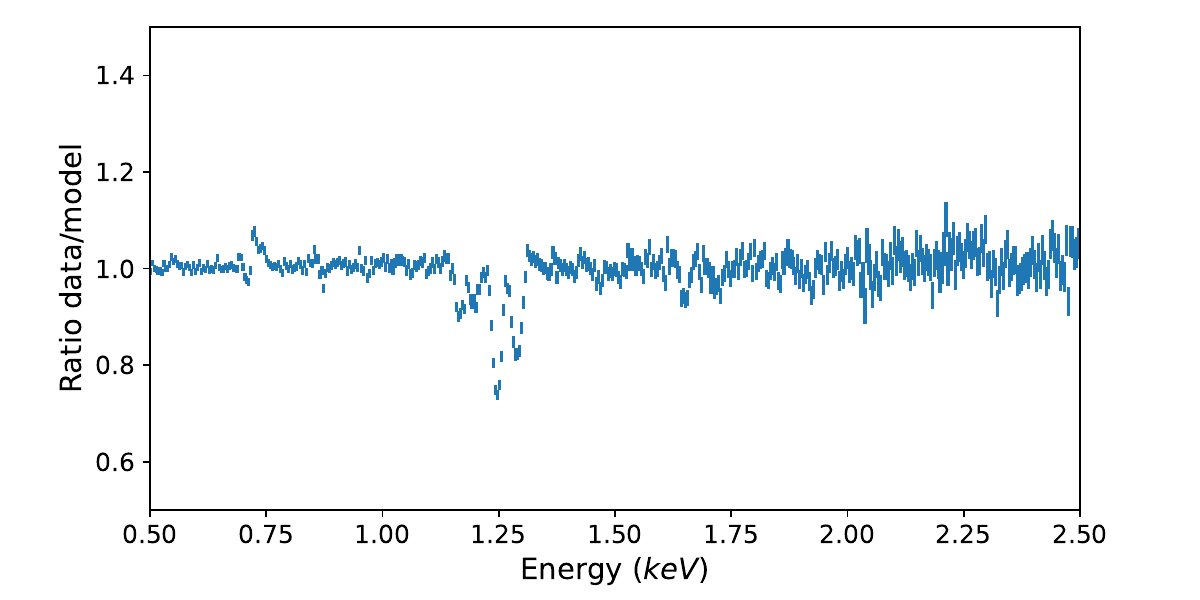}
\caption{A 100 ks simulated X-IFU spectrum for a wind with $log(\xi)=4, N_H=10^{23}, v_0=0.10c$ and a NLSy1 incident SED (blue points), best-fitted with a powerlaw-SED WINE table (red line). Bottom panel reports the data to best fit model ratio.}
\label{athena-spec}
\end{figure}

\begin{figure}
\centering
\includegraphics[width=\columnwidth]{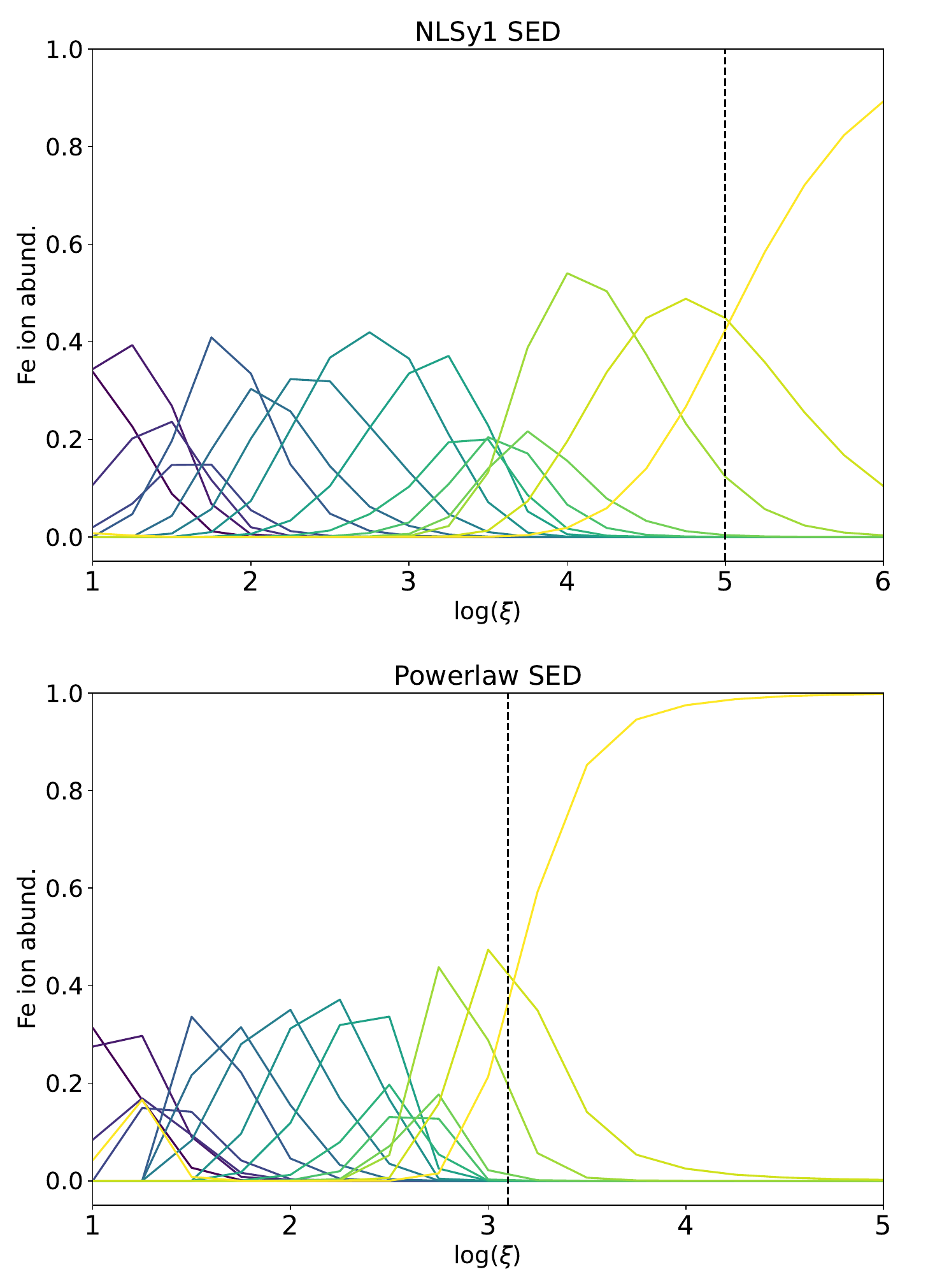}
\caption{Iron ion abundances (computed with XSTAR) as a function of $log(\xi)$ for an optically-thin slab and assuming either a NLSy1 (top) or a powerlaw (bottom) incident SED. Dark to light colours correspond to increasing ionic charge, with the lightest (and rightmost) line being Fe XXVII. Dashed vertical lines in the top and bottom panels correspond respectively to the input and the best-fit values (see text).}
\label{fe-abunds}
\end{figure}

\section{\textit{Athena X-IFU} simulations}
\label{athena-sims}
To assess the capabilities of the planned X-IFU microcalorimeter onboard the \textit{Athena} satellite, we repeat the same procedure of Sect. \ref{izw-sim} using the updated response matrices following the 2023 mission reformulation\footnote{X-IFU files are available at \url{https://x-ifu.irap.omp.eu/resources/for-the-community}}. We use the same input model and fitting strategy of Sect. \ref{izw-sim}. Again, we verified that adopting a 5 eV energy binning and the C-statistic does not change the results. As shown in Fig. \ref{athena-fill}, thanks to its higher effective area and resolution, X-IFU will be able to constrain $log(\xi_0), v_{out}, N_H$ with uncertainties smaller by a factor 2, 20 and 5, respectively and $\theta_{out}$ with half the uncertainty for input $log(\xi_0) \leq 4.0$.

Fig. \ref{athena-spec} shows a simulated spectrum with input $\log(\xi)=4, N_H=10^{23}, v_0=0.10$ and the best fit obtained by replacing the NLSy1-SED wind components with those assuming the powerlaw-SED. As shown in Fig. \ref{fe-abunds} (see also \citealp{nicastro99b}), the different incident SEDs lead to different ionic populations as a function of $log(\xi)$. As a result, the best fit model is unable to correctly reproduce the Fe L-shell features at $E \approx 1 keV$, which are due to a mixture of different ions. Thanks to this, X-IFU will be able to discriminate the incident SED through the absorption (and, eventually, emission) features in the soft band. 

Conversely, it can be seen that the relative abundance of Fe XXV, XXVI, XXVII is similar for the two SEDs, albeit with a shift in the value of $log(\xi)$. Therefore, fitting a spectrum simulated with a NLSy1 SED with a higher $log(\xi)=5$, for which these ions are dominants, with a powerlaw-SED table yields an acceptable fit with $log(\xi)=3.1$. As shown in Figure \ref{fe-abunds} (vertical dashed lines), these values of $log(\xi)$ lead to the same abundance of Fe XXV, XXVI, XXVII with both SEDs.

\section{Discussion and conclusions}
\label{discuss}

\subsection{Comparison with other wind models}
WINE is built on XSTAR, which is a one-dimensional photoionisation code. Such limitation is addressed by post-processing the gas emissivity and convolving it with three-dimensional line profiles. Therefore, its underlying structure is intrinsically different from three-dimensional Monte-Carlo radiative transfer codes such as DISKWIND \citep{sim08,sim10a,sim10b,matzeu22} and MONACO \citep{odaka11,hagino17} (see \citealp{noebauer19} for a review on the models available in the literature). Such codes are able to reproduce with great accuracy the scattering of the radiation by the wind as a function of its geometry and velocity field. However, to cope with the huge computing time required for such simulations, they are based on a number of assumption concerning the kinematics and the geometry of the wind. Moreover, their photoionisation computation is optimised for a highly-ionised gas and focuses mainly on transitions from the K, L, M atomic shells.
Due to their fundamental differences, a direct comparison between these models and WINE is not straightforward, but we expect WINE to be more accurate for relatively low $C_f$ and/or low $N_H$, i.e., in all those cases in which the reflected spectrum is subdominant with respect to the transmitted one.

The higher degree of freedom in modelling the radial density and velocity profiles in WINE also allows to better constrain the radial thickness of the wind $\Delta r$ (see Eqs. \ref{dr-cases}-\ref{v_r}) which, in turn, can probe $r_0, n_0$ (i.e., the launching radius and the wind number density), as done for the UFO detected in the X-ray spectrum of the AGN PG1448+273 \citep{llt21}. Together with the covering factor $C_f$ (which can be derived through WINE once the emission geometry is constrained), such quantities are crucial to accurately compute $\Dot{M}_{out}, \Dot{E}_{out}$. In turn, $\Dot{M}_{out}, \Dot{E}_{out}$ are fundamental to probe the impact of the wind on the surrounding environment. By comparing them with the energetics of galaxy-scale outflows, typically observed at millimetre to optical wavelengths, it is possible to determine the efficiency of the propagation from the nuclear to the host galaxy scales and, thus, to assess the impact on the galaxy life cycle and on the AGN fuelling (the so-called 'AGN feedback'; see e.g.  \citealp{Fiore17,cicone18,marasco20,tozzi21}).

We finally note that the column density slicing performed in WINE for the ionisation and radiative transfer is the same of the MHD wind model of \citealp{fkc10,fukumura18,fukumura22}. We aim to extend their analysis of magnetic disc winds through absorption features with the inclusion of the emission spectrum. Moreover, the velocity, density and, thus, ionisation profiles predicted for radiatively-driven winds, such as in DISKWIND or in \cite{lne21}, can be reproduced in WINE, therefore allowing for the comparison of different launching mechanisms within the same code.

\begin{figure*}
\centering
\includegraphics[width=1.8\columnwidth]{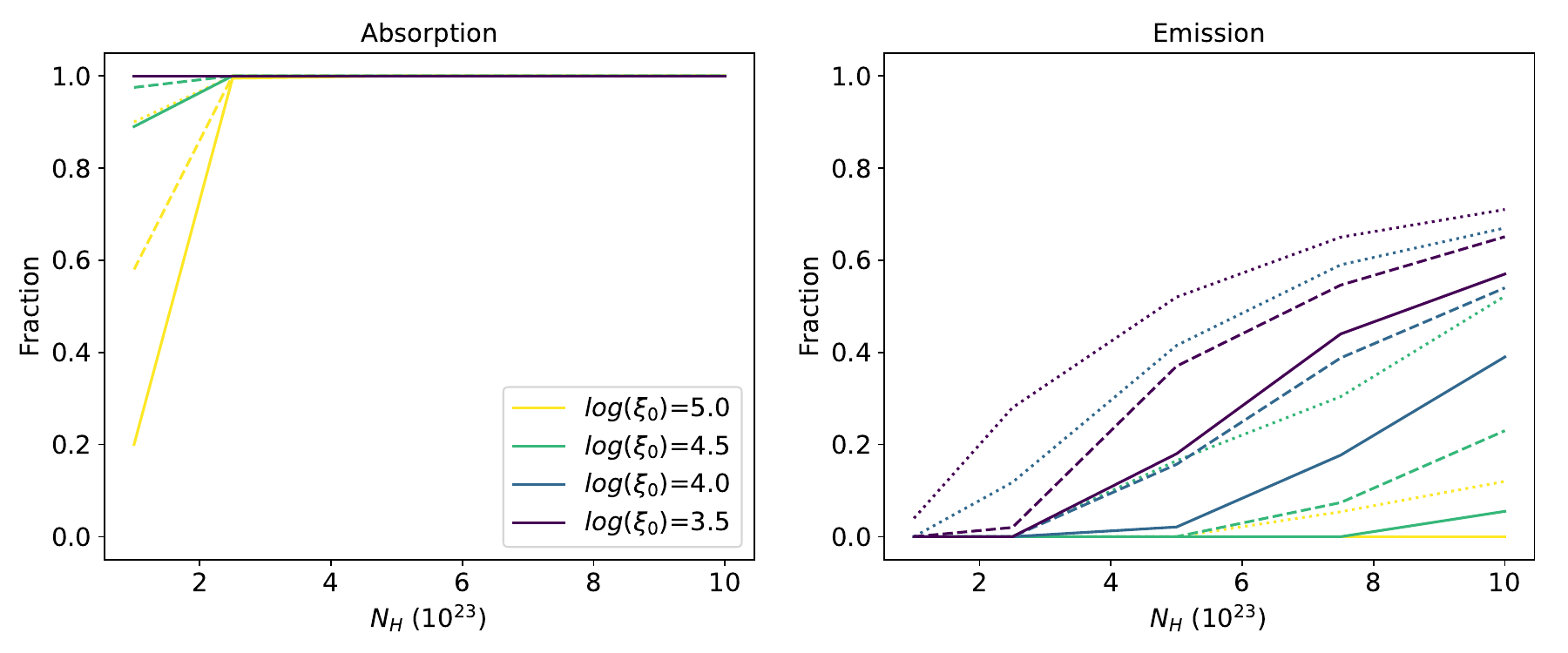}
\caption{Fraction of the \textit{XRISM Resolve} simulations in which the WINE absorption (left) and emission (right) components are detected above a given statistical significance, as a function of the input $N_H$. Dotted, dashed and solid lines correspond to 2, 3, 4 $\sigma$, respectively.}
\label{significance}
\end{figure*}

\begin{figure}
\centering
\includegraphics[width=0.9\columnwidth]{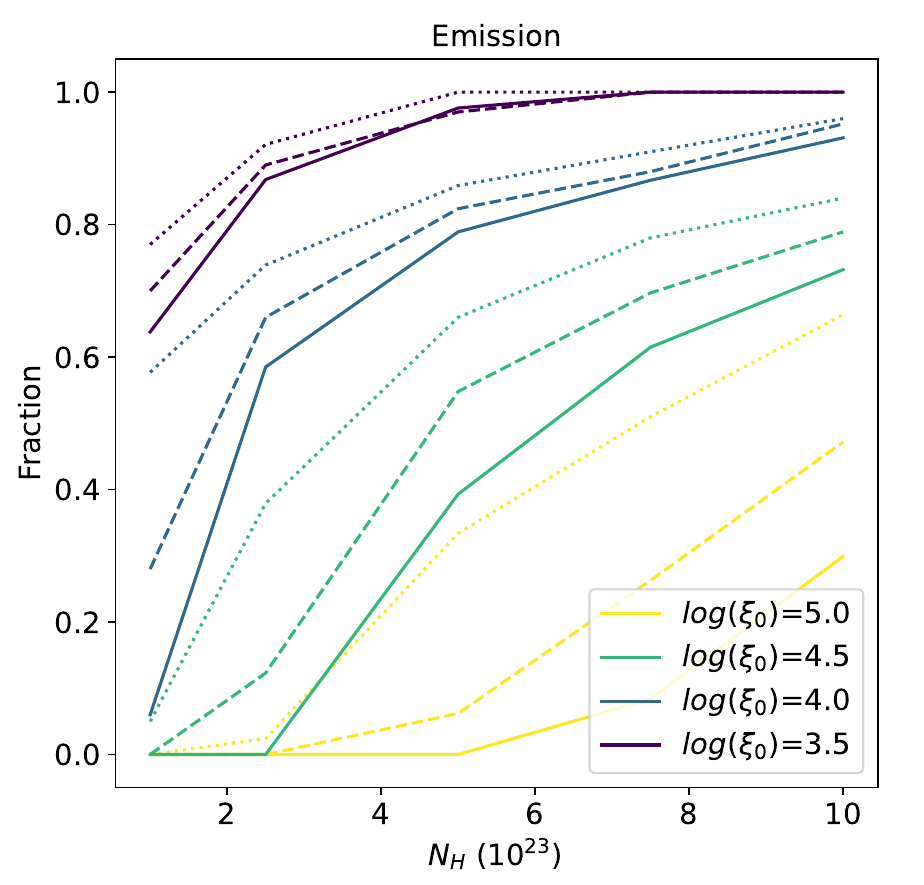}
\caption{Fraction of the \textit{Athena X-IFU} simulations in which the WINE emission is detected above a given statistical significance, as a function of the input $N_H$. Dotted, dashed and solid lines correspond to 2, 3, 4 $\sigma$, respectively.}
\label{athena-significance}
\end{figure}

\subsection{Summary of the results}
In this paper we outlined the design, initial conditions and physical picture of the Wind in the Ionised Nuclear Environment (WINE) model. We illustrated some of its main results by simulating the disc winds frequently observed in the X-ray spectra of Active Galactic Nuclei. More precisely, we focused on Narrow-Line Seyfert 1 (NLSy1)-like galaxies, i.e. highly-accreting AGNs, where most of the Ultra Fast Outflows (UFOs) for which both emission and absorption lines have been detected so far. Their combination gives rise to the so-called P-Cygni profiles, which are valuable spectroscopic signatures to investigate some of the most elusive wind properties (first of all, the covering factor $C_f$) and to shed light on the wind launching, which is hard to constrain by relying on absorption spectroscopy alone. 
Our main results are the followings:
\begin{itemize}
\item The gas opacity, kinematics and geometry all play a pivotal role in shaping the emission spectrum. The gas opacity depends on the ionisation and on the total (Hydrogen-equivalent) column density, which together dictate the ionic column giving rise to the different atomic transitions. Kinematics, i.e, the outflow and turbulent velocities, and the geometry regulate the broadening of the emission profiles via special relativistic Doppler effects. All of these properties must be accounted for to correctly model the emission profiles and, in turn, to derive meaningful physical constraints on the outflowing gas.
\item The incident SED has a strong impact on the gas ionisation. For a given $\xi$, the gas is much more ionised in the case of a powerlaw SED than in the case of a softer NLSy1- like SED. The gas opacity and, thus, the strength of the wind absorption and emission features are considerably reduced, as shown in Fig. \ref{sedcomparison}. This may be one of the reason why P-Cygni profiles are more frequently detected in NLSy1-like sources. 

A similar trend has been observed for UV winds as well. Specifically, the velocity of winds traced by both blueshifted CIV emission and blueshifted, broad CIV absorption correlates with the steepness of the ionising spectrum, usually parametrised through the $\alpha_{OX}$ ratio\footnote{The $\alpha_{OX}$ is defined as the slope connecting the monochromatic luminosities at 2500 Angstrom and 2 keV.} \citep{nardini19,rankine20,vietri20,vietri22,saccheo23}.

\item We assessed the power of the \textit{Resolve} microcalorimeter onboard {\it XRISM} in constraining the UFO parameters - including the emission - for a source with an X-ray flux typical of the targets of the Science Performance Verification phase ($F_{2-10 keV}=10^{-11} erg\ cm^{-2} s^{-1}$, see Sect. \ref{resolve-sim} for details). Assuming an input $v_{out}=0.1 c$ we find that, as shown in Fig. \ref{fig_simul} \textit{Resolve} will be able to constrain $log(\xi_0), v_{out}$ with $<5\%$ accuracy and $N_H$ with $<40\%$ accuracy for a wind with input $log(\xi_0) \leq 4.0$. For a wind with input $log(\xi_0) \geq 4.5$ and $N_H \geq 5 \cdot 10^{23}$, the accuracy of the best fit values is within 6\% for $log(\xi_0), v_{out}$ and 40\% for $N_H$. The opening angle of the emitting region \tout\ can be determined within a $\approx$60\% accuracy, provided that the input \tout\ is $\geq 45 \deg$. When considering any input \tout\, the accuracy reduces to a factor 2.5 for input $log(\xi_0) \leq 4.0, N_H \geq 5 \cdot 10^{23}$ (see Appendix \ref{xrism-sims-appendix} for further details). Fig. \ref{significance} reports, as a function of input $N_H$, the fraction of simulations in which the absorption and the emission components (left and right panels, respectively) are detected with a significance higher than 2, 3, 4 $\sigma$. For the absorption, the significance is computed as the difference between the best fit $\chi^2$ obtained by fitting the simulated data with a simple powerlaw continuum and with a powerlaw modified by the WINE absorption ($WINE_{abs} \cdot powerlaw$). For the emission, the significance is estimated from the difference between the previous absorption-only model and the full model including the wind emission ($WINE_{abs} \cdot powerlaw+ WINE_{em}$). While the absorption is detected with $\geq 4 \sigma$ significance for any input $\xi_0$ and $N_H \geq 2.5 \cdot 10^{23}$, the significance of the emission strongly depends on the wind opacity (i.e., it increases for increasing $N_H$ and decreasing $\xi_0$).

The constraining power is the same for different velocities, provided that the observed opacity is the same, i.e. that the intrinsic $N_H$ is increased(decreased) for increasing(decreasing) velocity, to compensate for the special relativistic reduction of the opacity according to Eq. \ref{eq:abs-rel}. The same trend also holds for the \textit{X-IFU} simulations discussed below.

\item Thanks to its higher effective area and resolution, the \textit{X-IFU} microcalorimeter onboard \textit{Athena} will allow to constrain the wind properties with even significantly higher accuracy. The absorption is significantly detected in 100\% of the simulations, for every input $log(\xi_0)$ and $N_H$. As shown in Fig. \ref{athena-significance}, the emission is detected with much higher significance than with \textit{XRISM Resolve} (see Fig. \ref{significance}). For input $\log(\xi_0)=3.5, 4.0$ and $N_H \geq 5 \cdot 10^{23} cm^{-2}$, in $>$75\% of the simulations the emission is detected at 4$\sigma$ significance, while for input $\log(\xi_0)=4.5, N_H \geq 5 \cdot 10^{23} cm^{-2}$, 50\% of the simulations are above the 3$\sigma$ threshold.

Moreover, the Fe L-shell features around 1 keV, which are due to a mixture of different ions, will also allow to discriminate between a soft and a steep incident ionising SED. While this is not currently possible with \textit{Resolve} onboard \textit{XRISM}, since the Gate Valve limits the sensitivity to $E \gtrsim 2 keV$, most of its targets will likely be already known, therefore allowing one to reconstruct their SED through data from other observatories.
\end{itemize}

\subsection{Future upgrades}
We plan three main areas of improvement for WINE. First, the computing time needed to build large table models (currently of the order of tens of days for a 50 CPU server) can be dramatically reduced by adopting neural network emulators, such as that presented in \cite{matzeu22}. Then, we will increase the modularity of WINE by giving the user the possibility to compute ionisation balance and radiative transfer with other publicly available codes, such as Cloudy \citep{Cloudy17}. This will allow to test the impact of the assumptions, limitations and atomic databases of the different codes on the resulting spectra and to take advantage of the most recent updates of the different codes.

Finally, we aim to implement in WINE an updated version of the Time Evolving PhotoIonisation Device (TEPID; \citealp{luminari23b}) which will be able to model both the time-variable gas absorption and emission spectra, thanks to the inclusion of a detailed electron level population treatment. TEPID will allow to extend wind modelling beyond the assumption of photoionisation equilibrium limit. Indeed, time-evolving ionisation is particularly relevant in AGNs, where the ionising luminosity is observed to vary dramatically. The gas equilibration timescale (i.e., the typical time needed by the gas to adjust to luminosity variations) is inversely proportional to its number density $n$ \citep{nicastro99} and, for the typical values of X-ray disc winds (especially for Warm Absorbers) it can be longer than the luminosity variation timescale, leading to a non-equilibrium ionisation dynamic  \citep{rogantini22,sadaula22,luminari23b,gu23}.

We also note that the perturbation-free assumption in WINE can be relaxed in two main ways. First, it is possible to modify the density and velocity profiles, Eqs. \ref{WINE:n_r}, \ref{WINE:v_r}, introducing a stochastic fluctuation pattern at perturbation length scales on top of the "ideal" powerlaw trend. Secondly, the synthetic spectra of a multiphase wind can be represented through the superposition of the absorption/emission features due to the different outflow phases. 
We will also consider to introduce further degrees of freedom in WINE to account for the wind anisotropy, e.g. through the density contrast $\delta \rho / \langle \rho \rangle$.

\subsection{Public release of WINE table models}
We make publicly available the WINE table models presented in Sect. \ref{resolve-sim} at \url{https://baltig.infn.it/ionisation/wine}. They can be implemented within the most popular fitting tools, such as \textit{xspec}, \textit{sherpa} \citep{sherpa} and \textit{spex} \citep{spex}. The two tables, one with the soft, NLSy1-like SED and one with the powerlaw SEDs, cover a parameter range suited for X-ray disc winds in AGNs, as well as in compact accreting sources, such as X-ray Binaries \citep{fks17,tominaga23} and Ultraluminous X-ray Sources \citep{pinto16,kosec21}.

\begin{acknowledgements}
AL thanks Massimo Cappi for useful discussions. All the authors thank the IT office from the Roma2 INFN section, particularly Federico Zani, and the \textit{rmlab} infrastructure for providing computing power for the WINE simulations and for assisting in the creation of the website. We also thank Diego Paris, Elena De Rossi and Federico Fiordoliva from the Astronmical Observatory of Rome (INAF) for computational support and Giustina Vietri for discussions on UV outflows. AL, EP, FN acknowledge support from the HORIZON-2020 grant “Integrated Activities for the High Energy Astrophysics Domain" (AHEAD-2020), G.A. 871158). EP, FT acknowledges funding from the European Union - Next Generation EU, PRIN/MUR 2022 2022K9N5B4. We used {\sc Astropy},\footnote{\url{http://www.astropy.org}} a community-developed core {\sc Python} package for Astronomy \citep{astropy18}, {\sc numpy} \citep{numpy} and {\sc matplotlib} \citep{matplotlib}. 
\end{acknowledgements}

\bibliographystyle{aa}
\bibliography{sbs}

\begin{appendix}
\section{Perpendicular velocity components}
\label{appendix-vrot}
The main effect of an eventual velocity component perpendicular to the line of sight, such as that due to Keplerian rotation, will be to increase the angle $\theta$ between the total wind velocity and the line of sight (LOS; which for a pure outward radial motion is =0) and, thus, to modify the relativistic factors $\psi, \psi^3$ of Eq. \ref{en_proj}, regulating respectively the wavelength shift and the beaming in the wind reference frame. To assess the impact of this on the wind interaction with the radiation field, we compared $\psi, \psi^3$ for a pure outward motion (as assumed in WINE) with $v_{out}=0.1, 0.3$ c and adding a transverse rotational component corresponding to the Keplerian rotational velocity v$_{rot}=\sqrt{\frac{2 G M_{BH}}{r_0}}$ for different launching radii $r_0$, i.e. $\psi_{rot}, \psi_{rot}^3$.

\begin{figure}
\centering
\includegraphics[width=\columnwidth]{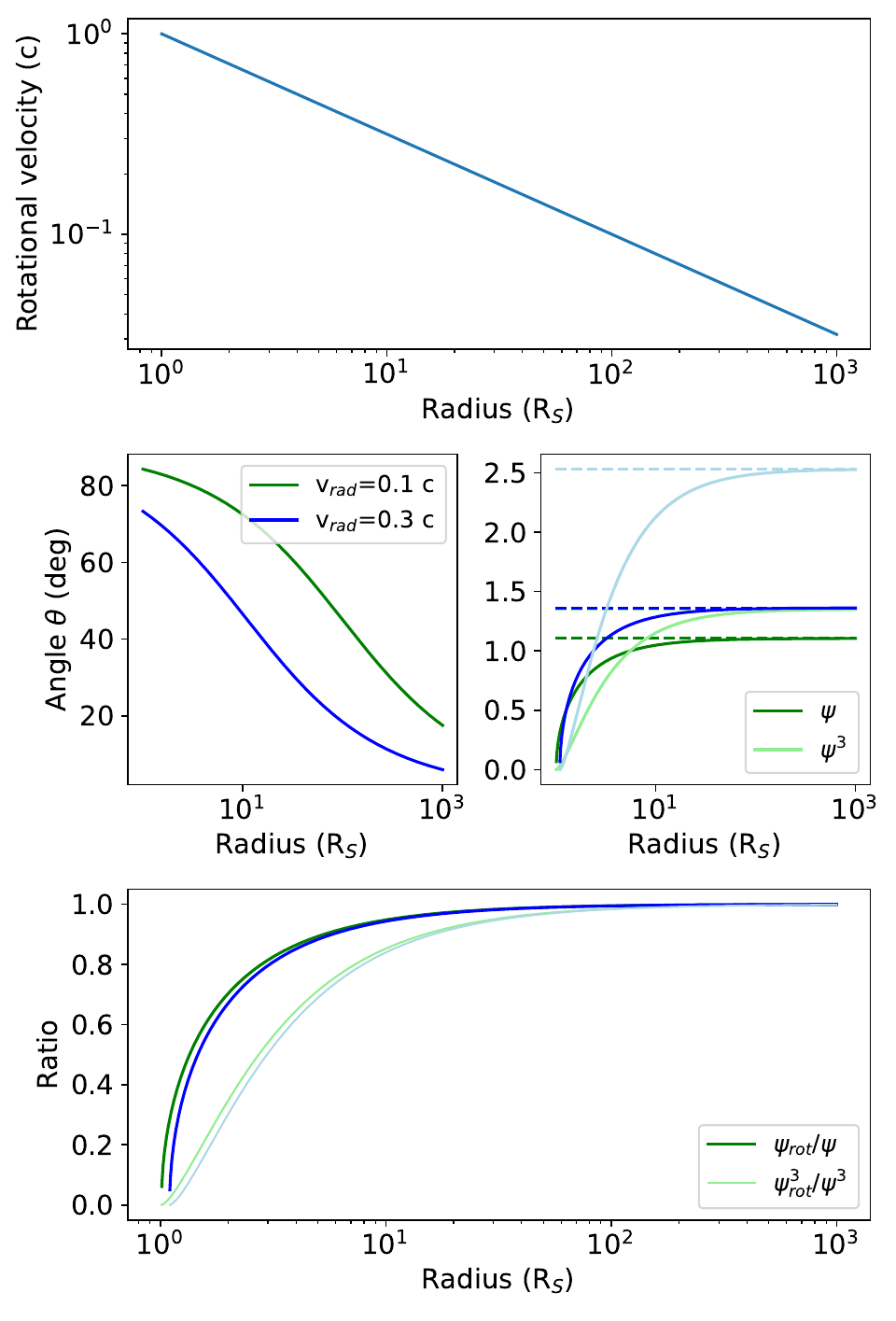}
\caption{Impact of a velocity component perpendicular to the line of sight. Top: rotational velocity $v_{rot}$ (in units of $c$) as a function of the disc radius (in units of $R_S$). Middle, both as a function of the radius; left: angle $\theta$ between the line of sight and the total velocity for $v_{rad}=$0.1, 0.3 c (colour coding; see legend); right: relativistic factors $\psi_{rot}, \psi^3_{rot}$ for the same total velocity (dark and light colours, respectively). For comparison, dashed lines report $\psi, \psi^3$, i.e. the same factors computed neglecting $v_{rot}$. Bottom: ratio between the relativistic factors: dark ad light colours correspond to $\psi_{rot}/\psi$ and $\psi^3_{rot}/\psi^3$, respectively. Colour coding as above.}
\label{fig:vrot}
\end{figure}

Fig. \ref{fig:vrot}. top panel, shows v$_{rot}$ as a function of $r_0$. Middle panels report, as a function of $r_0$, $\theta$ (left) and $\psi_{rot}, \psi_{rot}^3$ (right) including v$_{rot}$ and v$_{out}$=0.1, 0.3 c. For comparison, $\psi, \psi^3$ are plotted with dashed lines. Within $\sim$ 10 R$_S$, the inclusion of v$_{rot}$ leads to a strong increase of $\theta$, i.e. the motion is almost perpendicular to the LOS. Consequently, both $\psi_{rot}, \psi^3_{rot}$ are very small, resulting in spectroscopic signatures with small blueshift and limited interaction between the gas and the radiation field, respectively. For increasing r$_0$, v$_{rot}$ decreases and, accordingly, $\theta$ is closer to 0 and $\psi_{rot}, \psi^3_{rot}$ tend to $\psi, \psi^3$. The bottom panel reports the ratios $\psi_{rot}/\psi$ and $\psi^3_{rot}/\psi^3$. For r$_0$=10 R$_S$, $\psi_{rot}/\psi=$0.93 and $\psi^3_{rot}/\psi^3=0.85$, while for r$_0$=100 R$_S$ they are both $>$0.999. The blueshift and the opacity of the observed absorption features as measured by WINE (and, thus, the derived v and $N_H$) can be corrected by multiplying them by $\psi_{rot}/\psi$ and $\psi^3_{rot}/\psi^3$, provided that the radial location of the wind (or, equivalently, the ratio between outflow and rotational velocity components) can be estimated. Such a-posteriori correction cannot be performed for the emission profiles which, unlike the absorption, are integrated over the outflow solid angle. However, the ratios above provide an overview of the expected variation of the velocity broadening and equivalent width of the profiles, respectively.

\section{Comparison between WINE and XSTAR}
\begin{figure}
\centering
\includegraphics[width=\columnwidth]{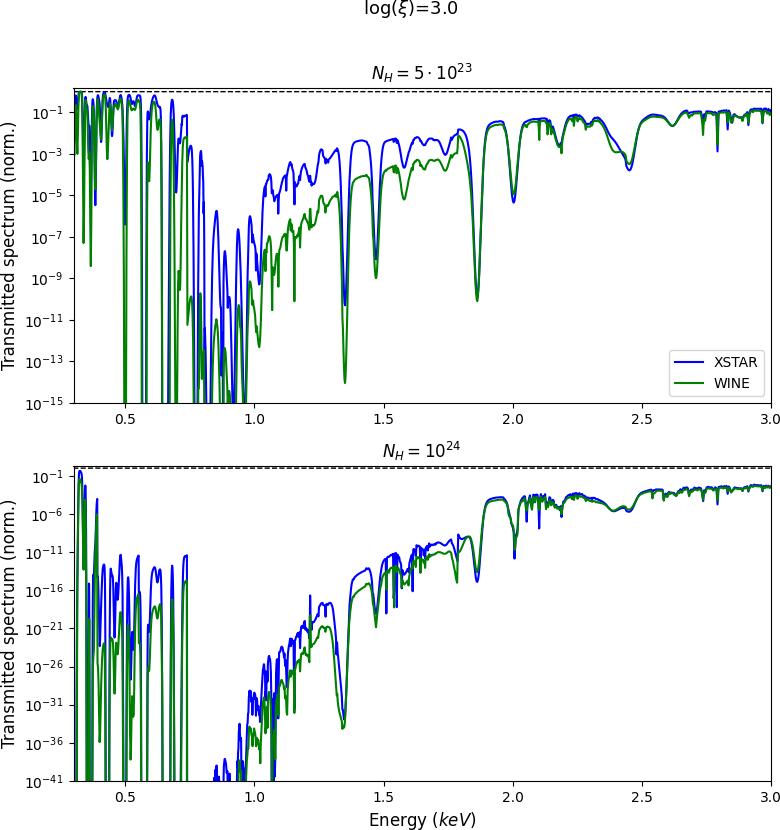}
\caption{Comparison between WINE- and XSTAR-generated absorption spectra for $log(\xi)=3$ and $N_H=5 \cdot 10^{23}, 10^{24}$ (top and bottom panel, respectively). Blue lines correspond to the XSTAR results, green to the WINE ones. For ease of visualisation, the spectra are divided by the incident continuum.}
\label{fig:xstar-comp}
\end{figure}

Fig. \ref{fig:xstar-comp} shows the comparison between the absorption spectra computed with WINE and XSTAR (green and blue lines, respectively) for $\xi=10^3$ and $N_H=5 \cdot 10^{23}, 10^{24}$ (top and bottom panel, respectively). As discussed in Sect. \ref{scheme}, WINE spectra are built by dividing the wind column in a series of thin slabs (with a linear thickness of $5 \cdot 10^{22}$ in this example) and computing photoionisation and radiative transfer iteratively, from the innermost to the outermost. Wind emission is computed in post-processing, therefore the emitted radiation does not affect radiative transfer and the resulting absorption spectrum. Conversely, XSTAR spectra are computed giving as input the whole column density and, thus, the emitted radiation is accounted for automatically by the code during the radiative transfer. This leads to slightly smaller opacities in the XSTAR spectra. However, as can be seen in the Figure, the differences in terms of fluxes per unit keV are very small, of the order of $10^{-5}$ and $10^{-16}$ with respect to the incident continuum for $N_H=5 \cdot 10^{23}$ and $10^{24}$, respectively, showing that the WINE slicing does not lead to dramatic changes in the absorption spectra even for such high wind opacities.

\section{\textit{XRISM Resolve} simulations}
\label{xrism-sims-appendix}
\begin{figure}
\centering
\includegraphics[width=\columnwidth]{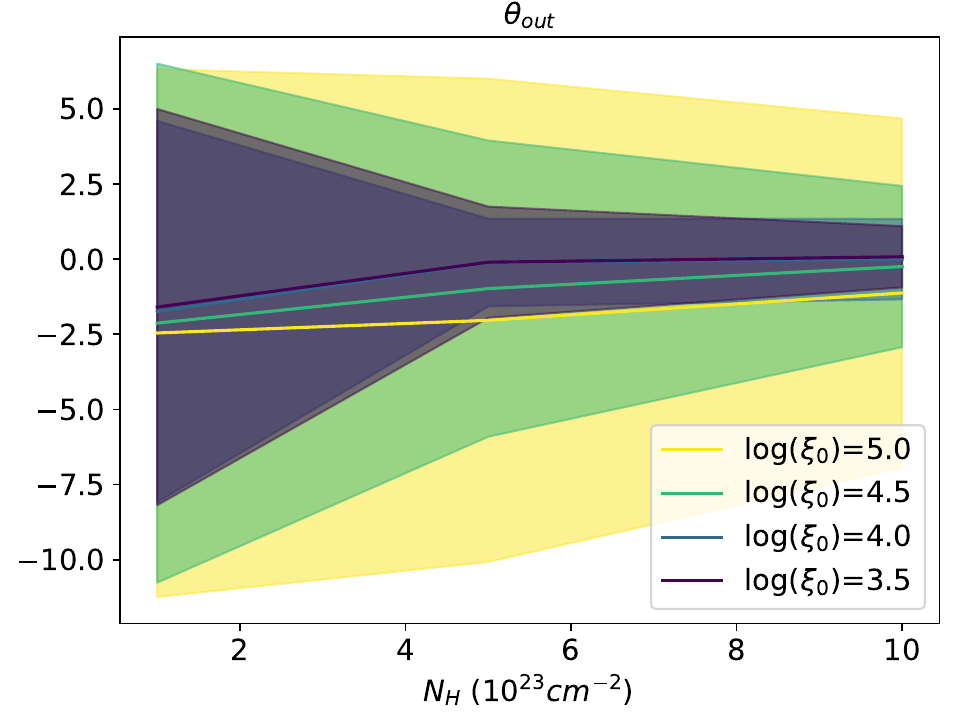}
\caption{Ratio $\Delta$ and corresponding standard deviations for \textit{XRISM Resolve} simulations, relative to \tout, as a function of input $N_H$ and for different input $log(\xi_0)$ (see legend).}
\label{fig:xrism-tout}
\end{figure}

Fig. \ref{fig:xrism-tout} shows the ratio $\Delta$ between input and best fit values for \tout, including all the simulations.

\section{\textit{Athena X-IFU} simulations}
\begin{figure*}
\centering
\includegraphics[width=1.9\columnwidth]{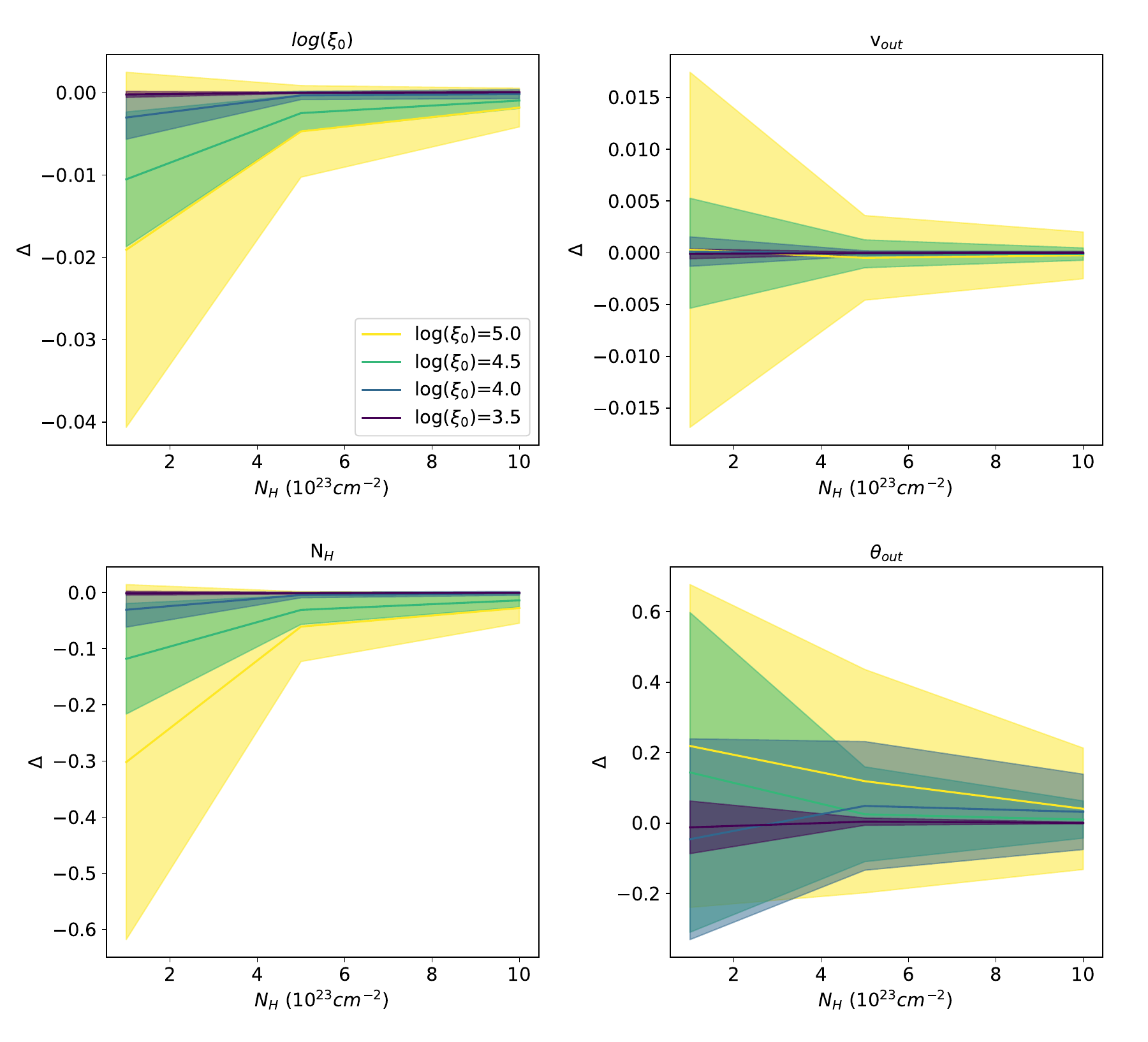}
\caption{Ratio $\Delta$ and corresponding standard deviations for \textit{Athena X-IFU}, as a function of input $N_H$ and for different input $log(\xi_0)$ (see legend).}
\label{athena-fill}
\end{figure*}

Figure \ref{athena-fill} shows the ratio $\Delta$ for the \textit{Athena} X-IFU simulations. For ease of comparison with the \textit{XRISM} results in Fig. \ref{fig_simul}, bottom right panel reports only the cases with input $\theta_{out}>45 deg$.

\end{appendix}

\end{document}